\documentclass[amsmath, pra, twocolumn, showpacs,aps,10pt]{revtex4-1}
\usepackage{graphics}
\usepackage{longtable}
\begin{document}
%
%

\title{Dynamical polarizability of atoms in arbitrary light fields: general theory and application to cesium}
 
\author{Fam Le Kien}
\altaffiliation{Also at Institute of Physics, Vietnamese Academy of Science and Technology, Hanoi, Vietnam.}

\author{P. Schneeweiss}

\author{A. Rauschenbeutel} 

\affiliation{Vienna Center for Quantum Science and Technology, Institute of Atomic and Subatomic Physics, Vienna University of Technology, Stadionallee 2, 1020 Vienna, Austria}

\date{\today}

\begin{abstract}
We present a systematic derivation of the dynamical polarizability and the ac Stark shift of the ground and excited states of atoms interacting with a far-off-resonance light field  of arbitrary polarization. We calculate the scalar, vector, and tensor polarizabilities of atomic cesium using resonance wavelengths and reduced matrix elements for a large number of transitions. We analyze the properties of the fictitious magnetic field produced by the vector polarizability in conjunction with the ellipticity of the polarization of the light field. 
\end{abstract}

\maketitle

\section{Introduction} 
\label{sec:introduction}

One of the main motivations of current laser cooling and trapping techniques is to use atoms
for storing and processing quantum
information that is encoded in the atomic states by means of resonant or near-resonant light.
Due to the weak coupling of neutral atoms to their environment,
coherent manipulation of atomic states can be robust against external
perturbations  \cite{Metcalf99}. This makes optically trapped neutral atoms prime candidates for, e.g., the
implementation of quantum memories and quantum repeaters \cite{Kimble08,Polzik10,Molmer10}.
For atom trapping, far-off-resonance laser fields are used because they ensure low
scattering rates, compatible with long coherence times.
The presence of these intense far-detuned light fields shifts the energy levels of the atom. In general, the light shift (ac Stark shift) depends not only on the dynamical polarizability of the atomic state and on the light intensity but also on the polarization of the field. For this reason, various experimental situations require a systematic study of the dynamical polarizability of the ground and excited states of atoms interacting with a far-off-resonance light field  of arbitrary polarization. In particular, this becomes important for optical trapping using near-fields or nonparaxial light beams.
One example is nanofiber-based atom traps, which have recently been realized  \cite{Rauschenbeutel10,Goban12} and in which the 
nanofiber-guided trapping light fields are evanescent waves in the fiber transverse plane \cite{fibermode}.
Another example is tightly focused optical dipole traps, where
the longitudinal polarization component of a nonparaxial light beam can lead to significant internal-state decoherence \cite{Regal12,Vuletic12}.
Plasmonically enhanced optical fields \cite{Chang09,Gullans12} also have, in general, complex local polarizations. 
Therefore, the calculation of the resulting optical potentials in all these cases requires a suitable formalism to take polarization effects into account.

Despite a large number of works on the polarizabilities of atoms, most of the previous calculations were devoted to the static limit \cite{Schmieder,Khadjavi,Safronova99,Safronova04,Clark10}. 
Accurate polarizabilities for a number of atoms of the periodic table have been calculated by a
variety of techniques \cite{Clark10}. These include the sum-over-states method, which is based on the use of available experimental and/or theoretical data, and the direct methods, which are based on \textit{ab initio} calculations of atomic wave functions. 
The \textit{ab initio} calculations of atomic structures involve the refined many-body perturbation theory, the relativistic
coupled-cluster calculations, or the random phase method \cite{Clark10}.
High-precision \textit{ab initio} calculations of atomic polarizability have been performed using the relativistic
all-order method in which all single, double, and partial triple excitations of the Dirac-Fock wave functions are
included to all orders of perturbation theory  \cite{Safronova99,Safronova04}.
Recently, in order to search for magic wavelengths \cite{Katori99} for a far-off-resonance trap, the dynamical scalar and tensor polarizabilities as well as the light shifts of the ground and excited states of strontium \cite{Katori99,Katori05} and cesium \cite{McKeever03,Hakuta05} have been calculated for a wide range of light wavelengths. The principal idea of magic wavelengths is based on a clever choice of the trapping light wavelength for which the excited  and ground states of an atom experience shifts of equal sign and magnitude  \cite{Katori99}. Magic wavelengths have been found for atomic cesium in red-detuned traps ~\cite{Katori99,McKeever03} and in combined two-color (red- and blue-detuned) traps \cite{Hakuta05}. Searches for magic and tune-out wavelengths of a number of alkali-metal atoms (from Na to Cs) have been conducted by calculating dynamical polarizabilities using a relativistic coupled-cluster method \cite{Arora07,Arora11}. All the three components of the dynamical polarizability, that is, the scalar, vector, and tensor polarizabilities \cite{Manakov86}, and the associated ac Stark shifts  have been calculated for the cesium clock states \cite{Rosenbusch09,Mabuchi06,Jessen10}. 
Calculations of the adiabatic potentials for atomic cesium in far-off-resonance nanofiber-based traps \cite{Rauschenbeutel10,Dowling96,twocolors,Vetsch10} have been performed \cite{Rauschenbeutel10,Hakuta05,twocolors,Vetsch10,Lacroute12}. The vector polarizability was omitted in \cite{Hakuta05,twocolors}, but was included in the calculations for the ac Stark shifts in Ref. \cite{Lacroute12}. The scalar, vector, and tensor polarizabilities of atomic rubidium have recently been calculated \cite{Arora12}.  

Due to the complexity of the calculations for the dynamical polarizability of a realistic multilevel atom, various approximations have been used and different expressions for the components of the dynamical polarizability have been presented in different treatments. One example is that the counter-rotating terms in the atom--field interaction Hamiltonian was neglected in  Refs. \cite{Mabuchi06,Jessen10} but was taken into account in Refs. \cite{Hakuta05,Arora07,Arora11,Manakov86,Rosenbusch09}. Another example is that the definition for the reduced matrix element used in Refs. \cite{Hakuta05,Arora07,Arora11,Manakov86,Rosenbusch09} is different from that in Refs. \cite{Mabuchi06,Jessen10,Lacroute12}. Furthermore, the coupling between different hyperfine-structure (hfs) levels of the same fine-structure state was taken into account in Refs. \cite{Hakuta05,Arora07} but was neglected in Refs. \cite{Rosenbusch09,Mabuchi06,Jessen10}. In addition, the numerical calculations require the use of resonance wavelengths and reduced matrix elements of a large number of  atomic transitions, which are not available in a single source. Since the authors of previous works often did not describe in detail the formalisms and the data they used, it is not easy to see the connections between their results and to employ them correctly. 

The purpose of this article is to provide a systematic treatment of the dynamical polarizability of the ground and excited states of atoms interacting with a far-off-resonance light field of arbitrary polarization. We specify all theoretical definitions and tools necessary for computing the light shifts of atomic levels. Based on the approach of Rosenbusch \textit{et al.}~\cite{Rosenbusch09}, we provide the details of the derivation of the expressions for the ac Stark interaction operator and for the scalar, vector, and tensor components of the dynamical polarizability. 
We also discuss the light-induced fictitious magnetic field.
We supply a comprehensive set of experimental and theoretical data for resonance wavelengths and reduced matrix elements for a large number of atomic transitions that allows one to perform the computation of the light shifts of the levels associated with the $D_2$-line transition of cesium. 
Furthermore, we present the results of numerical calculations for the corresponding components of the polarizability for a wide range of light wavelengths.
Both, the atomic data and the numerical results are provided as electronic files which accompany this article \cite{Supplementary}.

\section{ac Stark shift and atomic polarizability} 
\label{sec:theory}

In this section, we present the basic expressions for the ac Stark shift operator and the scalar, vector, and tensor polarizabilities of a multilevel atom interacting with a far-off-resonance light field of arbitrary polarization  \cite{Manakov86,Rosenbusch09,Mabuchi06,Jessen10}. We also provide the results of numerical calculations for atomic cesium for a wide range of light wavelengths.

\subsection{General theory}
\label{subsection:theory}

\subsubsection{Hyperfine interaction}

\begin{figure}[tbh]
\begin{center}
  \includegraphics{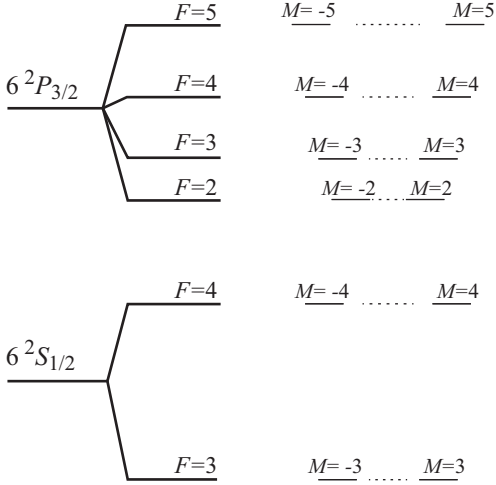}
 \end{center}
\caption{Energy levels associated with the $D_2$ line of a cesium atom.}
\label{fig1}
\end{figure}

We consider a multilevel atom.
We use an arbitrary Cartesian coordinate frame $\{x,y,z\}$, with $z$ being the quantization axis. In this coordinate frame, we specify bare basis states of the atom (see Fig. \ref{fig1} for the levels associated with the $D_2$-line transition of cesium). 
Due to the hfs interaction, the total electronic angular momentum $\mathbf{J}$ is coupled to the nuclear spin $\mathbf{I}$. The hfs interaction is described by the operator \cite{Metcalf99}
\begin{equation}
V^{\mathrm{hfs}}=\hbar A_{\mathrm{hfs}}\,\mathbf{I}\cdot\mathbf{J}+\hbar B_{\mathrm{hfs}}\frac{6(\mathbf{I}\cdot\mathbf{J})^2+3\mathbf{I}\cdot\mathbf{J}-2\mathbf{I}^2\mathbf{J}^2}{2I(2I-1)2J(2J-1)}.
\label{n2}
\end{equation}
Here, $A_{\mathrm{hfs}}$ and $B_{\mathrm{hfs}}$ are the hfs constants. 
Note that  $A_{\mathrm{hfs}}$ and $B_{\mathrm{hfs}}$ depend on the fine-structure level $|nJ\rangle$. In the case of atomic cesium, the values of these constants are $A_{\mathrm{hfs}}/2\pi=2298.1579425$ MHz \cite{Arimondo77} and $B_{\mathrm{hfs}}/2\pi=0$ for the ground state $6S_{1/2}$ and $A_{\mathrm{hfs}}/2\pi=50.28827$ MHz and $B_{\mathrm{hfs}}/2\pi=-0.4934$ MHz \cite{Tanner03} for the excited state $6P_{3/2}$. We also note that high-order hfs interaction effects, which mix different fine-structure levels $|nJ\rangle$, have been omitted in expression (\ref{n2}) for the hfs interaction operator $V^{\mathrm{hfs}}$.

Due to the hfs interaction, the projection $J_z$ of the total electronic angular momentum $\mathbf{J}$ onto the quantization axis $z$ is not conserved. However, in the absence of the external light field, the projection $F_z$ of the total angular momentum of the atom, described by the operator $\mathbf{F}=\mathbf{J}+\mathbf{I}$, onto the quantization axis $z$ is conserved. We use the notation $|nJFM\rangle$ for the atomic hfs basis ($F$ basis) states, where $F$ is the quantum number for the total angular momentum $\mathbf{F}$ of the atom, 
$M$ is the quantum number for the projection $F_{z}$ of $\mathbf{F}$ onto the quantization axis $z$, 
$J$ is the quantum number for the total angular momentum $\mathbf{J}$ of the electron,
and $n$ is the set of the remaining quantum numbers $\{nLSI\}$, with $L$ and $S$ being the quantum numbers for the total orbital angular momentum and the total spin of the electrons, respectively.
In the hfs basis $\{|nJFM\rangle\}$, the operator $V^{\mathrm{hfs}}$ is diagonal. The nonzero matrix elements of this operator are  
\begin{eqnarray}
\lefteqn{\langle nJFM|V^{\mathrm{hfs}}|nJFM\rangle=\frac{1}{2}\hbar A_{\mathrm{hfs}}G} 
\nonumber\\&&\mbox{}
+\hbar B_{\mathrm{hfs}}\frac{\frac{3}{2}G(G+1)-2I(I+1)J(J+1)}{2I(2I-1)2J(2J-1)}, 
\label{n3}
\end{eqnarray}
where $G=F(F+1)-I(I+1)-J(J+1)$.

\subsubsection{ac Stark interaction}

Consider the interaction of the atom with a classical light field
\begin{equation}
\mathbf{E}=\frac{1}{2}\boldsymbol{\mathcal{E}}e^{-i\omega t}+\mathrm{c.c.}
=\frac{1}{2}\mathcal{E}\mathbf{u}e^{-i\omega t}+\mathrm{c.c.},
\label{n1}
\end{equation}
where $\omega$ is the angular frequency and $\boldsymbol{\mathcal{E}}=\mathcal{E}\mathbf{u}$ is the positive-frequency electric field envelope,
with $\mathcal{E}$ and $\mathbf{u}$ being the field amplitude and the polarization vector, respectively.
In general, $\mathcal{E}$ is a complex scalar and $\mathbf{u}$ is a complex unit vector. 

We assume that the light field is far from resonance with the atom. In addition, we assume that $J$ is a good quantum number. This means that we treat only the cases where the Stark interaction energy is small compared to the fine structure splitting.
In the dipole approximation, the interaction between the light field and the atom can be described by the operator
\begin{equation}\label{m1}
V^{E}=-\mathbf{E}\cdot\mathbf{d}=-\frac{1}{2}\mathcal{E}\mathbf{u}\cdot\mathbf{d}e^{-i\omega t}
-\frac{1}{2}\mathcal{E}^*\mathbf{u}^*\cdot\mathbf{d}e^{i\omega t},
\end{equation}
where $\mathbf{d}$ is the operator for the electric dipole of the atom. 
When the light field is far from resonance with the atom, the second-order ac Stark shift of a nondegenerate atomic energy level $|a\rangle$ is, as shown in Appendix \ref{sec:two-level atom}, given by \cite{Manakov86,Rosenbusch09,Boyd}
\begin{eqnarray}\label{m2}
\delta E_a&=&-\frac{|\mathcal{E}|^2}{4\hbar}\sum_b\mathrm{Re}
\bigg(\frac{|\langle b|\mathbf{u}\cdot\mathbf{d}|a\rangle|^2}{\omega_b-\omega_a-\omega-i\gamma_{ba}/2}
\nonumber\\&&\mbox{}
+\frac{|\langle a|\mathbf{u}\cdot\mathbf{d}|b\rangle|^2}{\omega_b-\omega_a+\omega+i\gamma_{ba}/2}\bigg).
\end{eqnarray}
Here, $|a\rangle$ and $|b\rangle$ are the atomic eigenstates with unperturbed energies $\hbar\omega_a$ and $\hbar\omega_b$, respectively, and with spontaneous decay rates $\gamma_a$ and $\gamma_b$, respectively, while $\gamma_{ba}=\gamma_a+\gamma_b$ is the transition linewidth. 
We can consider the energy shift (\ref{m2}) as an expectation value $\delta E_a=\langle a|V^{EE}|a\rangle$, where
\begin{equation}\label{m3}
V^{EE}=\frac{|\mathcal{E}|^2}{4}[(\mathbf{u}^*\cdot\mathbf{d}) \mathcal{R}_+(\mathbf{u}\cdot\mathbf{d})
+(\mathbf{u}\cdot\mathbf{d}) \mathcal{R}_-(\mathbf{u}^*\cdot\mathbf{d})],
\end{equation}
with
\begin{eqnarray}\label{m4}
\mathcal{R}_+&=&-\frac{1}{\hbar}\sum_b \mathrm{Re} \left(\frac{1}{\omega_b-\omega_a-\omega-i\gamma_{ba}/2}\right) |b\rangle\langle b|,
\nonumber\\
\mathcal{R}_-&=&-\frac{1}{\hbar}\sum_b \mathrm{Re}\left(\frac{1}{\omega_b-\omega_a+\omega+i\gamma_{ba}/2}\right) |b\rangle\langle b|.
\end{eqnarray}
We assume that $V^{EE}$ is the operator for the ac Stark interaction \cite{Manakov86,Rosenbusch09}, i.e., that it correctly describes not only the level shift but also the level mixing of nondegenerate as well as degenerate states. 
While this educated guess has not been derived from first principles,
it is consistent with the results of the second-order perturbation theory for the dc Stark shift \cite{Schmieder,Khadjavi} and of the Floquet formalism for the ac Stark shift  \cite{Manakov86,Rosenbusch09}.

\subsubsection{Atomic polarizability}

Let us examine the energy shifts of levels of a single fine-structure state $|nJ\rangle$. In general, due to the degeneracy of atomic levels and the possibility of level mixing, we must diagonalize the interaction Hamiltonian in order to find the energy level shifts. Since the atomic energy levels are perturbed by the Stark interaction and the hfs interaction, the combined interaction Hamiltonian is 
\begin{equation}
H_{\mathrm{int}}=V^{\mathrm{hfs}}+V^{EE}.
\label{n40}
\end{equation}
In terms of the hfs basis states $|(nJ)FM\rangle\equiv|nJFM\rangle$,
the Stark operator $V^{EE}$, given by Eq. (\ref{m3}), can be written as
\begin{equation}
V^{EE}=\sum_{FMF'M'}V_{FMF'M'}^{EE}   |(nJ)FM\rangle\langle (nJ)F'M'|,
\label{n10}
\end{equation}
where $V_{FMF'M'}^{EE}\equiv\langle (nJ)FM|V^{EE}|(nJ)F'M'\rangle$ are the matrix elements and
are given as \cite{Rosenbusch09}
\begin{eqnarray}
V_{FMF'M'}^{EE}&=&\frac{1}{4}|\mathcal{E}|^2
\sum_{^{K=0,1,2}_{q=-K,\dots,K}} \alpha^{(K)}_{nJ}
\{\mathbf{u}^*\otimes\mathbf{u}\}_{Kq}
\nonumber\\&&\mbox{}
\times(-1)^{J+I+K+q-M}
\sqrt{(2F+1)(2F'+1)}
\nonumber\\&&\mbox{}
\times\bigg(\begin{array}{ccc}
F &K &F' \\
M & q& -M'
\end{array}\bigg)
\bigg\{\begin{array}{ccc}
F &K &F' \\
J &I &J 
\end{array}\bigg\}.
\label{n11}
\end{eqnarray}
Here we have introduced the notations 
\begin{eqnarray}
\alpha^{(K)}_{nJ}
&=&(-1)^{K+J+1}\sqrt{2K+1}
\nonumber\\&&\mbox{}
\times\sum_{n'J'}(-1)^{J'}
\bigg\{\begin{array}{ccc}
1 &K &1 \\
J &J'&J 
\end{array}\bigg\}
|\langle{n' J'\|\mathbf{d}\|n J}\rangle|^2
\nonumber\\&&\mbox{}
\times \frac{1}{\hbar}\mathrm{Re}\bigg(\frac{1}{\omega_{n'J'nJ}
-\omega-i\gamma_{n'J'nJ}/2}
\nonumber\\&&\mbox{}
+\frac{(-1)^K}{\omega_{n'J'nJ}+\omega+i\gamma_{n'J'nJ}/2}\bigg),
\label{n6}
\end{eqnarray}
with $K=0, 1, 2$, for the reduced dynamical scalar ($K=0$), vector ($K=1$), and tensor ($K=2$) polarizabilities of the atom in the fine-structure level $|nJ\rangle$. 
In Eqs. (\ref{n11}) and (\ref{n6}), we have employed the notations $\bigl( \begin{smallmatrix} j_1&j_2&j\\ m_1& m_2&m\end{smallmatrix}\bigr)$ 
and $\bigl\{\begin{smallmatrix} j_1&j_2&j_3\\ j_4&j_5&j_6\end{smallmatrix}\bigr\}$ for the Wigner 3-$j$ and 6-$j$ symbols, respectively.
The notations $\omega_{n'J'nJ}=\omega_{n'J'}-\omega_{nJ}$ and $\gamma_{n'J'nJ}=\gamma_{n'J'}+\gamma_{nJ}$
stand for the angular frequency and linewidth, respectively, of the transition between the fine-structure levels $|n' J'\rangle$ and $|n J\rangle$. 
The details of the derivation of Eqs. (\ref{n11}) and (\ref{n6}) are given in Appendix \ref{sec:derivation}. 
Note that the above-defined polarizabilities are just the real parts of the complex polarizabilities. The imaginary parts of the complex polarizabilities are related to the scattering rate of the atom \cite{Jackson99}. 

The compound tensor components $\{\mathbf{u}^*\otimes\mathbf{u}\}_{Kq}$ in Eq. (\ref{n11}) are defined as
\begin{eqnarray}
\{\mathbf{u}^*\otimes\mathbf{u}\}_{Kq}&=&
\sum_{\mu,\mu'=0,\pm 1}
(-1)^{q+\mu'} u_{\mu} u^*_{-\mu'}
\nonumber\\&&\mbox{}\times
\sqrt{2K+1}\bigg(\begin{array}{ccc}
1 &K &1 \\
\mu &-q &\mu' 
\end{array}\bigg).
\label{n12}
\end{eqnarray}
Here, $u_{-1}=(u_{x}-iu_{y})/\sqrt{2}$, $u_0=u_{z}$, and $u_{1}=-(u_{x}+iu_{y})/\sqrt{2}$ are the spherical tensor components of the polarization vector $\mathbf{u}$ in the Cartesian coordinate frame $\{ x,y,z\}$. 

The reduced matrix elements $\langle n' J' \|\mathbf{d}\|n J\rangle$  of the electric dipole in Eq. (\ref{n6}) can 
be obtained from the oscillator strengths 
\begin{equation}\label{n8}
f_{nJn'J'}=\frac{2m_e\omega_{n'J'nJ}}{3\hbar e^2} \frac{1}{2J+1} |\langle n'J'\|\mathbf{d}\|nJ\rangle|^2,
\end{equation}
where $m_e$ is the mass of the electron and $e$ is the elementary charge,
or from the transition probability coefficients 
\begin{equation}\label{n9}
A_{n'J'nJ}=\frac{\omega_{n'J'nJ}^3}
{3\pi\epsilon_0\hbar c^3 }\frac{1}{2J'+1}|\langle n'J'\|\mathbf{d}\|nJ\rangle|^2.
\end{equation}

We note that the Stark interaction operator (\ref{n10}) with the matrix elements (\ref{n11}) can be written in the form
\cite{Rosenbusch09,Manakov86}
\begin{eqnarray}
\lefteqn{V^{EE}= -\frac{1}{4}|\mathcal{E}|^2 \bigg\{\alpha^s_{nJ} 
-i\alpha^v_{nJ} \frac{[\mathbf{u}^*\times\mathbf{u}]\cdot\mathbf{J}}{2J}}
\nonumber\\&&\mbox{}
+\alpha^T_{nJ} \frac{3[(\mathbf{u}^*\cdot\mathbf{J})(\mathbf{u}\cdot\mathbf{J})
+(\mathbf{u}\cdot\mathbf{J})(\mathbf{u}^*\cdot\mathbf{J})]
-2\mathbf{J}^2}{2J(2J-1)}\bigg\}. 
\label{n4}
\end{eqnarray}
Here, $\alpha^s_{nJ}$,  $\alpha^v_{nJ}$, and $\alpha^T_{nJ}$ are the conventional dynamical scalar, vector, and tensor polarizabilities, respectively, of the atom in the fine-structure level $|nJ\rangle$. They are given as \cite{Rosenbusch09} 
\begin{eqnarray}
\alpha^s_{nJ}&=&\frac{1}{\sqrt{3(2J+1)}} \alpha^{(0)}_{nJ},
\nonumber\\
\alpha^v_{nJ}&=&-\sqrt{\frac{2J}{(J+1)(2J+1)}}\alpha^{(1)}_{nJ},
\nonumber\\
\alpha^T_{nJ}&=&-\sqrt{\frac{2J(2J-1)}{3(J+1)(2J+1)(2J+3)}}\alpha^{(2)}_{nJ}.
\label{n5}
\end{eqnarray}

Note that for $J=1/2$ and $K=2$, the Wigner 6-$j$ symbol in Eq. (\ref{n6}) is zero. Thus, the tensor polarizability vanishes for $J=1/2$ states (e.g., the ground states of alkali-metal atoms). In the case of linearly polarized light, the polarization  vector $\mathbf{u}$ can be taken as a real vector. In this case, the vector product $[\mathbf{u}^*\times\mathbf{u}]$ vanishes, making the contribution of the vector polarizability to the ac Stark shift to be zero. We  also note that $\gamma_{n'J'nJ}$ can be omitted from the denominators  in Eqs. (\ref{m2}), (\ref{m4}), and (\ref{n6}) when the light field frequency $\omega$ is far from resonance with the atomic transition frequencies $\omega_{n'J'nJ}$.

In general, $V^{EE}$ is not diagonal neither in $F$ and nor in $M$. Therefore,
in order to find the new eigenstates and eigenvalues, one has to diagonalize the  
Hamiltonian (\ref{n40}), which includes both the hfs splitting and the ac Stark interaction.
However, in  the case where the Stark interaction energy is small compared to the hfs splitting, we can neglect the mixing of atomic energy levels with different quantum numbers $F$.
In this case, the Stark operator $V^{EE}$ for the atom in a particular hfs level $|nJF\rangle$ can be presented in the form \cite{Jessen10}
\begin{eqnarray}
\lefteqn{V^{EE}= -\frac{1}{4}|\mathcal{E}|^2 \bigg\{\alpha^s_{nJF} 
-i\alpha^{v}_{nJF} \frac{[\mathbf{u}^*\times\mathbf{u}]\cdot\mathbf{F}}{2F}}
\nonumber\\&&\mbox{}
+\alpha^{T}_{nJF}\frac{3[(\mathbf{u}^*\cdot\mathbf{F})(\mathbf{u}\cdot\mathbf{F})
+(\mathbf{u}\cdot\mathbf{F})(\mathbf{u}^*\cdot\mathbf{F})]
-2\mathbf{F}^2}{2F(2F-1)}\bigg\},\qquad
\label{n16}
\end{eqnarray}
where
\begin{eqnarray}
\alpha^s_{nJF}&=&\alpha^s_{nJ}=\frac{1}{\sqrt{3(2J+1)}} \alpha^{(0)}_{nJ},\nonumber\\
\alpha^v_{nJF}&=&(-1)^{J+I+F}\sqrt{\frac{2F(2F+1)}{F+1}} 
\bigg\{\begin{array}{ccc}
F &1 &F \\
J &I&J 
\end{array}\bigg\}\alpha^{(1)}_{nJ},\nonumber \\
\alpha^T_{nJF}&=&-(-1)^{J+I+F}\sqrt{\frac{2F(2F-1)(2F+1)}{3(F+1)(2F+3)}} 
\nonumber \\&&\mbox{}
\times\bigg\{\begin{array}{ccc}
F &2 &F \\
J &I&J 
\end{array}\bigg\}\alpha^{(2)}_{nJ}.
\label{n17}
\end{eqnarray}
The coefficients $\alpha^s_{nJF}$, $\alpha^v_{nJF}$ and $\alpha^T_{nJF}$  are the conventional scalar, vector, and tensor polarizabilities of the atom, respectively, in a particular hfs level.
Note that the scalar  polarizability $\alpha^s_{nJF}$ does not depend on $F$.
This statement holds true only in the framework of our formalism, where the hfs splitting is omitted in the expression
for the atomic transition frequency $\omega_{n'J'F'nJF}$ in the calculations for the atomic polarizability,
that is, where the approximation $\omega_{n'J'F'nJF}=\omega_{n'J'nJ}$ is used. 
We also note that, if energies including hfs splittings are used in the denominators in the perturbation expression (\ref{m2}), then the wave functions of the states
$|a\rangle$ and $|b\rangle$ in the numerators should also incorporate hfs corrections to all orders of perturbation theory \cite{Rosenbusch09,Dzuba10}.
We emphasize that Eq. (\ref{n16}) is valid only when the coupling between different hfs levels 
$|nJF\rangle$ is negligible. Thus, Eq. (\ref{n16}) is less rigorous than Eq. (\ref{n4}). 

Furthermore, we note that, when the off-diagonal coupling is much smaller than  
the Zeeman splittings produced by an external magnetic field  $\mathbf{B}$, 
the mixing of different Zeeman sublevels can be discarded. In this case, the ac Stark shift of a Zeeman sublevel $|FM\rangle$ (specified in the quantization coordinate frame $\{x,y,z\}$ with the axis $z$ parallel to the direction $z_B$ of the magnetic field $\mathbf{B}$) is given by
\begin{eqnarray}
\Delta E_{\rm ac}&=&V_{FMFM}^{EE}
=-\frac{1}{4}|\mathcal{E}|^2 \bigg[\alpha^s_{nJF} +
C\alpha^v_{nJF}\frac{M}{2F}
\nonumber\\&&\mbox{}
-D\alpha^T_{nJF}\frac{3M^2-F(F+1)}{2F(2F-1)}\bigg],
\label{n18}
\end{eqnarray}
where
\begin{eqnarray}
C&=&|u_{-1}|^2-|u_1|^2=2\mathrm{Im}\,(u^*_{x}u_{y}),
\nonumber\\ 
D&=&1-3|u_{0}|^2=1-3|u_{z}|^2.
\label{n19}
\end{eqnarray}
The coefficients $C$ and $D$ are determined by the polarization vector $\mathbf{u}$ of the light field at the position of the atom. Note that the parameter $C$, which characterizes the vector Stark shifts, depends on the ellipticity of the light field in the transverse plane $(x,y)$. This parameter achieves its maximal magnitude $|C|=1$ when the longitudinal component of the field is absent and the light field is circularly polarized in the plane $(x,y)$. We also note that the parameter $D$, which characterizes the tensor Stark shifts, vanishes when $|u_{z}|=1/\sqrt3$.

\subsubsection{Fictitious magnetic field}

It is clear from Eqs. (\ref{n4}) and (\ref{n16}) that the effect of the vector polarizability on the Stark shift is equivalent to that of a magnetic field with the induction vector \cite{Cohen-Tannoudji72,Cho97,Zielonkowski98,Park01,Park02,Skalla95,Yang08,Rosatzin90,Wing84,Ketterle92,Kobayashi09} 
\begin{equation}\label{n20} 
\mathbf{B}^{\mathrm{fict}}
=\frac{\alpha^v_{nJ}}{8\mu_Bg_{nJ}J} i[\boldsymbol{\mathcal{E}}^*\times\boldsymbol{\mathcal{E}}]
=\frac{\alpha^v_{nJF}}{8\mu_Bg_{nJF}F} i[\boldsymbol{\mathcal{E}}^*\times\boldsymbol{\mathcal{E}}].
\end{equation}
Here, $\mu_B$ is the Bohr magneton and $g_{nJ}$ and $g_{nJF}$ are the Land\'{e} factors for
the fine-structure level $|nJ\rangle$ and the hfs level $|nJF\rangle$, respectively.
The nonrelativistic value of the Land\'{e} factor $g_{nJ}$ is given by \cite{Metcalf99}
\begin{eqnarray}\label{n21a}
g_{nJ}&=&g_L\frac{J(J+1)+L(L+1)-S(S+1)}{2J(J+1)} 
\nonumber\\&&\mbox{}
+g_S\frac{J(J+1)+S(S+1)-L(L+1)}{2J(J+1)}.
\end{eqnarray} 
Here, $g_L=1$ and $g_S\simeq 2.0023193$ are the orbital and spin g-factors for the electron, respectively.
When the contribution of the nuclear magnetic moment is neglected,
the Land\'{e} factor $g_{nJF}$ is
\begin{equation}\label{n21}
g_{nJF}=g_{nJ}\frac{F(F+1)+J(J+1)-I(I+1)}{2F(F+1)}.
\end{equation}

The direction of the light-induced fictitious magnetic field $\mathbf{B}^{\mathrm{fict}}$ is determined by the vector 
$i[\boldsymbol{\mathcal{E}}^*\times\boldsymbol{\mathcal{E}}]$, which is a real vector.
Similar to a real magnetic field, the fictitious magnetic field $\mathbf{B}^{\mathrm{fict}}$ is a pseudovector, that is,
$\mathbf{B}^{\mathrm{fict}}$ does not flip under space reflection. 
Another similarity is that both the real and fictitious magnetic fields flip under time reversal.
If the light field is linearly polarized, we have $i[\boldsymbol{\mathcal{E}}^*\times\boldsymbol{\mathcal{E}}]=0$ and hence $\mathbf{B}^{\mathrm{fict}}=0$. 
The middle expression in Eq. (\ref{n20}) shows that $\mathbf{B}^{\mathrm{fict}}$ is independent of $F$, that is, $\mathbf{B}^{\mathrm{fict}}$ is the same for all hfs levels $|nJF\rangle$ of a fine-structure level $|nJ\rangle$. Comparison between the middle and last expressions in Eq. (\ref{n20}) shows that
the factor $\alpha^v_{nJF}/g_{nJF}F$ does not depend on $F$. This conclusion is consistent with the relation
\begin{equation}\label{n17a}
\alpha^v_{nJF}= - \frac{F(F+1)+J(J+1)-I(I+1)}{(F+1)\sqrt{2J(J+1)(2J+1)}}
\alpha^{(1)}_{nJ},
\end{equation}
which can be obtained directly from the second expression in Eqs. (\ref{n17}) with the use of an explicit expression for the Wigner 6-$j$ symbol $\bigl\{\begin{smallmatrix} F&1&F\\ J&I&J\end{smallmatrix}\bigr\}$.

In general, the vector Stark shift operator can be expressed in terms of the operator $\mathbf{J}$ as
\begin{equation}{\label{n21b}} 
V^{EE}_{\mathrm{vec}}=\mu_Bg_{nJ}(\mathbf{J}\cdot\mathbf{B}^{\mathrm{fict}}).
\end{equation}
In the special case where the mixing of different hfs levels is negligible, that is, when $F$ is a good quantum number, the vector Stark shift operator can be expressed in terms of the operator $\mathbf{F}$ as 
\begin{equation}{\label{n22}} 
V^{EE}_{\mathrm{vec}}=\mu_Bg_{nJF}(\mathbf{F}\cdot\mathbf{B}^{\mathrm{fict}}).
\end{equation}
The vector form of Eqs. (\ref{n21b}) and (\ref{n22}) allows us to conclude that the fictitious magnetic field $\mathbf{B}^{\mathrm{fict}}$ can be simply added to a real static magnetic field $\mathbf{B}$ if the latter is present in the system [see Eqs. (\ref{n42}) and (\ref{n45}) in Appendix \ref{sec:magnetic}].

Let us discuss the case of the ground state $nS_{1/2}$ of an alkali-metal atom. In this case, we have $J=1/2$ and, therefore, $\alpha^T_{nJ}=0$. We assume that the hfs splitting of the ground state is very large compared to the Stark interaction energy. Then, the mixing of two different hfs levels $F=I\pm1/2$ of the ground state can be neglected, that is, $F$ can be considered as a good quantum number. It is obvious that $M$ is also a good quantum number when the quantization axis $z$ coincides with the direction of the fictitious magnetic field $\mathbf{B}^{\mathrm{fict}}$.

For the hfs levels $F=I\pm1/2$ of the ground state $nS_{1/2}$, we have $g_{nJF}|_{F=I+1/2}=-g_{nJF}|_{F=I-1/2}=g_{nJ}/(2I+1)$. 
When the hfs splitting of the ground state is very large compared to the light shift, the vector Stark shift operator is given in terms of the operator $\mathbf{F}$ by Eq. (\ref{n22}).
Hence, when the direction of the fictitious magnetic field $\mathbf{B}^{\mathrm{fict}}$ is taken as the quantization axis $z$, the vector Stark shifts of
the sublevels $M$ of the hfs levels $F=I+1/2$ and $F=I-1/2$ of the ground state are
\begin{equation}{\label{n22a}} 
V^{EE}_{\mathrm{vec}}|_{F=I+1/2}=\frac{\mu_Bg_{nJ}}{2I+1}MB^{\mathrm{fict}}
\end{equation}
and 
\begin{equation}{\label{n22b}} 
V^{EE}_{\mathrm{vec}}|_{F=I-1/2}=-\frac{\mu_Bg_{nJ}}{2I+1}MB^{\mathrm{fict}},
\end{equation}
respectively. These shifts are integer multiples of the quantity $\mu_Bg_{nJ}B^{\mathrm{fict}}/(2I+1)$. In other words, 
as expected from analogy with the well-known Zeeman effect,
the shifts are equidistant with respect to the quantum number $M$. 
It is clear that the sublevels $M$ and $-M$ of the hfs levels $F=I+1/2$ and $F=I-1/2$, respectively, of the ground state have the same vector Stark shift.
In contrast, the sublevels with the same number $M$ of two different hfs levels $F=I\pm1/2$ have opposite vector Stark shifts. Since the scalar Stark shift does not depend on $F$, the differential shift of the energies of the sublevels $M'$ and $M$ of the hfs levels $F'=I+1/2$ and $F=I-1/2$, respectively, of the ground state is just the differential vector Stark shift and is given by
\begin{eqnarray}
\Delta W_{M'M}&=&\frac{\mu_Bg_{nJ}}{2I+1}(M'+M)B^{\mathrm{fict}}
\nonumber\\
&=&\frac{\alpha^v_{nJ}}{8J(2I+1)}(M'+M) |i[\boldsymbol{\mathcal{E}}^*\times\boldsymbol{\mathcal{E}}]|.
\label{n47}
\end{eqnarray}
This differential shift vanishes when $M'+M=0$. This result is valid only in the framework of our formalism where the hfs splitting is neglected in the calculations for the atomic polarizability.

\subsection{Numerical calculations}
\label{subsection:calculations}

We now present the results of numerical calculations for the dynamical scalar, vector, and  tensor polarizabilities of the ground and excited states associated with the $D_2$-line transition of atomic cesium. Before we proceed, we note that, in order to search for red- and blue-detuned magic wavelengths for a far-off-resonance trap, the scalar and tensor polarizabilities of the ground and excited states of atomic cesium have been calculated \cite{McKeever03,Hakuta05,Arora07}. Relevant parameters were taken from a number of sources \cite{Safronova04,Fabry,Moore,Theodosiou}. Very recently, the vector light shifts of cesium atoms in a nanofiber-based trap have been studied \cite{Lacroute12}. However, the results for the vector polarizability have not been explicitly provided. 

Our calculations for the polarizabilities of cesium are based on Eqs. (\ref{n5}) in conjunction with Eqs. (\ref{n6}).
The calculations for the polarizability of the ground state $6S_{1/2}$ incorporate the couplings 
$6S_{1/2}\leftrightarrow (6\text{--}40)P_{1/2,3/2}$. The calculations for the polarizability of the excited state $6P_{3/2}$ incorporate the couplings $6P_{3/2}\leftrightarrow (6\text{--}40)S_{1/2}$ and $6P_{3/2}\leftrightarrow (5\text{--}42)D_{3/2,5/2}$. 
The energies of the levels with the principal quantum number $n\le 25$ are taken from \cite{NIST}. 
The energies of the levels with the principal quantum number $n\ge 26$ are provided by Arora and Sahoo \cite{Bindiya}. The reduced matrix elements for the transitions $6S_{1/2}\leftrightarrow (6\text{--}15)P_{1/2,3/2}$ are taken from \cite{Sieradzan}. 
The reduced matrix elements for the transitions $6P_{3/2}\leftrightarrow (6\text{--}10)S_{1/2}$ and $6P_{3/2}\leftrightarrow (5\text{--}8)D_{3/2,5/2}$ are taken from \cite{Arora07}. The reduced matrix elements for transitions to highly excited states are provided by Arora and Sahoo \cite{Bindiya}. These data were calculated  by using the relativistic all-order method, which includes single and double excitations  \cite{Arora07,Sieradzan}. The calculations for cesium were done in the same way as for rubidium \cite{Arora12}. 
The full set of parameters we used in our numerical calculations is given in Appendix \ref{sec:data}.
The states whose energy differences from the ground state are larger than the cesium ionization energy of 31406 cm$^{-1}$ provide a discrete representation of the continuum, similar to the calculations of Ref. \cite{Johnson08} for lithium. We add the contribution of the core, equal to $15.8$ a.u., to the results for the scalar polarizabilities \cite{Arora07}.
The polarizabilities are given in the atomic unit (a.u.) $e^2a_0^2/E_h$, where $a_0$ is the Bohr radius and $E_h=m_e e^4/(4\pi\epsilon_0\hbar)^2$ is the Hartree energy.

\begin{figure}[tbh]
\begin{center}
  \includegraphics{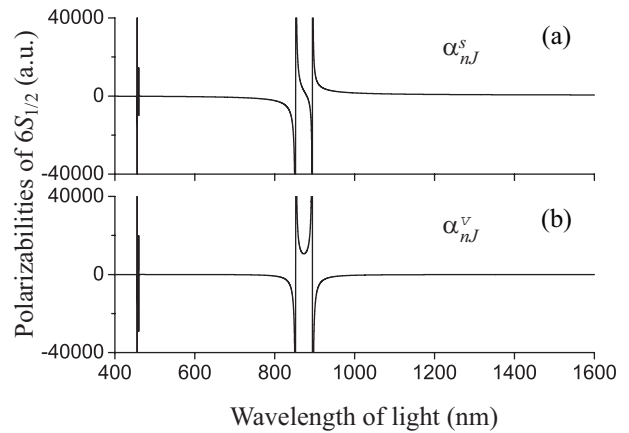}
 \end{center}
\caption{Scalar (a) and vector (b) polarizabilities $\alpha^s_{nJ}$ and $\alpha^v_{nJ}$, respectively,  of the
ground state $6S_{1/2}$ of atomic cesium as functions of the light wavelength $\lambda$. The data of this figure is provided as electronic files in \cite{Supplementary}.}
\label{fig2}
\end{figure}

We plot in Fig. \ref{fig2} the scalar and vector polarizabilities $\alpha^s_{nJ}$ and $\alpha^v_{nJ}$, respectively, of the ground state $6S_{1/2}$. As can be seen, in the region of wavelengths from 400 nm to 1600 nm, the profiles of both $\alpha^s_{nJ}$ and $\alpha^v_{nJ}$ have  two pairs of closely positioned resonances. 
One pair corresponds to the transitions between the ground state $6S_{1/2}$ and the excited state $6P_{1/2}$ ($D_1$ line, wavelength 894 nm) and the excited state $6P_{3/2}$ ($D_2$ line, wavelength 852 nm). The other pair corresponds to the transitions between the ground state $6S_{1/2}$ and the excited state $7P_{1/2}$ (wavelength 459 nm) and the excited state $7P_{3/2}$ (wavelength 455 nm). The effects of the other transitions are not substantial in this wavelength region. 
We note that our numerical calculations give the values $\alpha^s_{nJ}(6S_{1/2})\simeq 398.9$ a.u. and $\alpha^v_{nJ}(6S_{1/2})= 0$
for the scalar and vector polarizabilities, respectively, of the ground state $6S_{1/2}$ of atomic cesium in the static limit ($\omega=0$).
The static value $\alpha^s_{nJ}(6S_{1/2})\simeq 398.9$ a.u. is in agreement with the high-precision \textit{ab initio} theoretical values 
of $399.8$ a.u. \cite{Safronova99} and $398.2$ a.u. \cite{Safronova04} and the experimental value of $401$ a.u. \cite{Amini03}. 

\begin{figure}[tbh]
\begin{center}
  \includegraphics{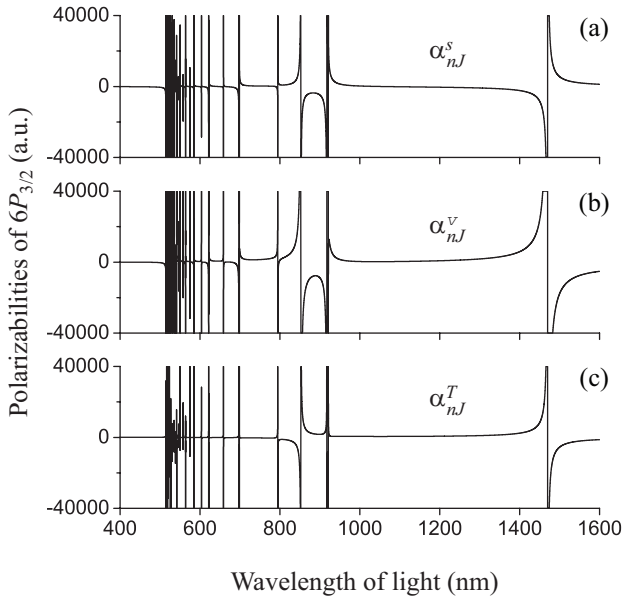}
 \end{center}
\caption{Scalar (a), vector (b), and tensor (c) polarizabilities $\alpha^s_{nJ}$, $\alpha^v_{nJ}$, and $\alpha^T_{nJ}$,  respectively, of the
excited state $6P_{3/2}$ of atomic cesium as functions of the light wavelength $\lambda$. The data of this figure is provided as electronic files in \cite{Supplementary}.}
\label{fig3}
\end{figure}

We plot in Fig. \ref{fig3} the scalar, vector, and tensor polarizabilities $\alpha^s_{nJ}$, $\alpha^v_{nJ}$, and $\alpha^T_{nJ}$,  respectively,  of the excited state $6P_{3/2}$. The figure shows that all the three components have multiple resonances. The most dominant resonances are due to the transitions from  $6P_{3/2}$ to (6--8)$S_{1/2}$ and (5--8)$D_{3/2,5/2}$.
We note that our numerical calculations give the values $\alpha^s_{nJ}(6P_{3/2})\simeq 1639.6$ a.u., $\alpha^v_{nJ}(6P_{3/2})= 0$,
and $\alpha^T_{nJ}(6P_{3/2})\simeq -260.4$ a.u. for the scalar, vector, and tensor polarizabilities, respectively, of the excited state $6P_{3/2}$ of atomic cesium in the static limit ($\omega=0$).
The static values $\alpha^s_{nJ}(6P_{3/2})\simeq 1639.6$ a.u. and $\alpha^T_{nJ}(6P_{3/2})\simeq -260.4$ a.u.
are in agreement with the high-precision \textit{ab initio} theoretical values of 
$1650$ a.u.  and $-261$ a.u. \cite{Arora07}, respectively, and with the experimental values of $1641$ a.u. and $-262$ a.u., respectively \cite{Tanner88}.

\begin{figure}[tbh]
\begin{center}
  \includegraphics{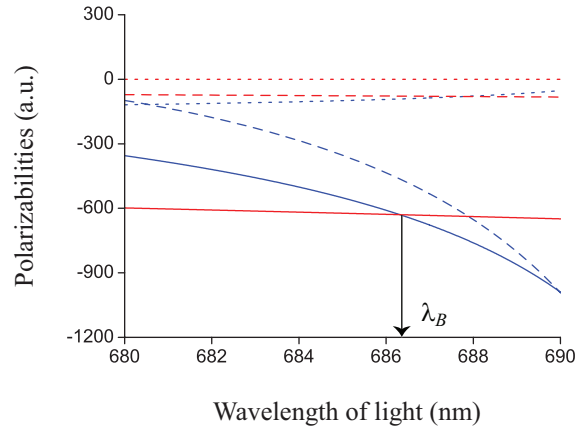}
 \end{center}
\caption{Polarizabilities of the ground state $6S_{1/2}$ (red color) and the excited state $6P_{3/2}$ (blue color) of atomic cesium in the region of blue-detuned wavelengths from 680 nm to 690 nm. The scalar, vector, and tensor components $\alpha^s_{nJ}$, $\alpha^v_{nJ}$, and  $\alpha^T_{nJ}$ are shown by the solid, dashed, and dotted curves, respectively.}
\label{fig4}
\end{figure}

\begin{figure}[tbh]
\begin{center}
  \includegraphics{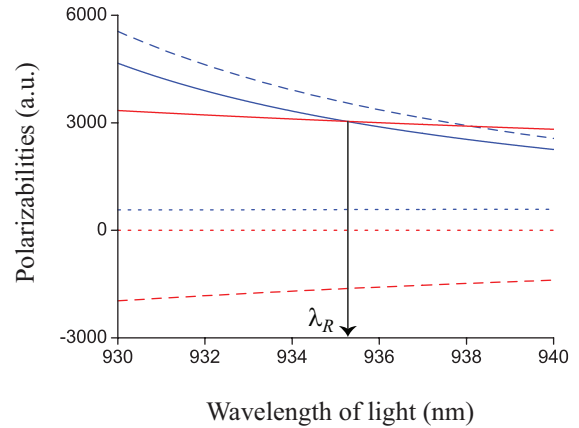}
 \end{center}
\caption{Same as Fig. \ref{fig4} but in the region of red-detuned wavelengths from 930 nm to 940 nm.}
\label{fig5}
\end{figure}

In order to display certain details, we plot in Figs. \ref{fig4} and \ref{fig5} the polarizabilities $\alpha^s_{nJ}$ (solid lines), $\alpha^v_{nJ}$ (dashed lines), and $\alpha^T_{nJ}$ (dotted lines) of the ground state $6S_{1/2}$ (red color) and the excited state  $6P_{3/2}$ (blue color) in two specific regions of wavelengths. Figures \ref{fig4} and \ref{fig5} show that the crossings of the scalar polarizabilities $\alpha^s_{nJ}(6S_{1/2})$ and $\alpha^s_{nJ}(6P_{3/2})$ of the ground and excited states, respectively, occur at the blue-detuned magic wavelength $\lambda_B\simeq 686.3$ nm \cite{Hakuta05} and the red-detuned magic wavelength $\lambda_R\simeq  935.2$ nm \cite{McKeever03}. Here, red and blue refer to the detunings with respect to the $D$-line transitions.
We observe from Figs.  \ref{fig4} and \ref{fig5} that the magnitude of the vector polarizability $\alpha^v_{nJ}$ is, in general, substantial compared to that of the scalar polarizability $\alpha^s_{nJ}$. Due to this fact, the vector polarizability can contribute significantly to the Stark shift when the polarization of the field is not linear. 

\begin{figure}[tbh]
\begin{center}
  \includegraphics{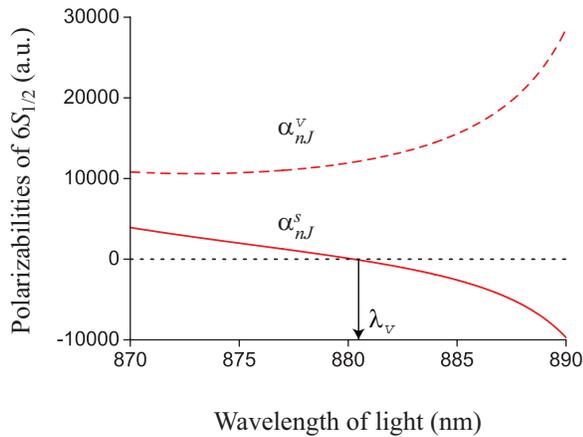}
 \end{center}
\caption{Scalar and vector polarizabilities $\alpha^s_{nJ}$ and $\alpha^v_{nJ}$, respectively,  of the
ground state $6S_{1/2}$ of atomic cesium for light wavelengths in the region from 870 nm to 890 nm.}
\label{fig6}
\end{figure}

Comparison between Figs. \ref{fig2}(a) and \ref{fig2}(b) shows that at the wavelength $\lambda_v\simeq 880.2$ nm, which lies between the $D_1$ and $D_2$ lines, the scalar polarizability $\alpha^s_{nJ}(6S_{1/2})$ of the ground state is vanishing while the vector polarizability $\alpha^v_{nJ}(6S_{1/2})$ of this state is significant (see Fig. \ref{fig6})  \cite{Cohen-Tannoudji72,Cho97,Zielonkowski98,Park01,Park02,Wing84,Rosatzin90,Ketterle92,Skalla95,Yang08,Kobayashi09}. At this specific wavelength, the ac Stark shifts of the sublevels of the atomic ground state are just the Zeeman-like shifts caused by a fictitious magnetic field $\mathbf{B}^{\mathrm{fict}}$. In other words, when specified in the quantization coordinate frame $\{x,y,z\}$ with the axis $z$ parallel to the direction of the vector product $i[\boldsymbol{\mathcal{E}}^*\times\boldsymbol{\mathcal{E}}]$, the sublevels $|FM\rangle$ of the ground state  will be shifted by an amount proportional to $(-1)^F Mi[\boldsymbol{\mathcal{E}}^*\times\boldsymbol{\mathcal{E}}]$. 

We note that the detunings $\Delta_2=\omega_{v}-\omega_{D_2}$ and $\Delta_1=\omega_{v}-\omega_{D_1}$ of the pure-vector-shift (scalar-shift-cancellation) frequency $\omega_{v}=2\pi c/\lambda_v$ from the $D_2$- and $D_1$-line transition frequencies $\omega_{D_2}$ and $\omega_{D_1}$, respectively, are such that  $\Delta_2/\Delta_1=-2.03\simeq -2$, in agreement with the results of Refs.  \cite{Cho97,Park01,Park02}. In order to understand this feature, we make a few additional approximations for the scalar polarizability $\alpha^{s}_{nJ}$ in the case where the level $|nJ\rangle$ is the ground state $nS_{1/2}$.
We keep only the excited levels $|n'J'\rangle=nP_{3/2}$ and $|n'J'\rangle=nP_{1/2}$ in the sum over $n'J'$ in Eq. (\ref{n6}). 
In the framework of the rotating-wave approximation, we neglect the counter-rotating term containing $\omega_{n'J'nJ}+\omega$
in Eq. (\ref{n6}). We also neglect $\gamma_{n'J'nJ}$ in the denominator of the co-rotating term containing $\omega_{n'J'nJ}-\omega$.
When we insert the result into the first expression in Eqs. (\ref{n5}), we obtain the following approximate expression for the scalar shift of the ground state:
\begin{equation}\label{n45b}
\alpha^{s}_{nJ}=-
\frac{|\langle{n P_{3/2}\|\mathbf{d}\|n S_{1/2}}\rangle|^2}{6\hbar\Delta_2}
-\frac{|\langle{n P_{1/2}\|\mathbf{d}\|n S_{1/2}}\rangle|^2}{6\hbar\Delta_1}.
\end{equation}
It is clear that $\alpha^{s}_{nJ}=0$ when 
\begin{equation}\label{n45c}
\frac{\Delta_2}{\Delta_1}=-\frac{|\langle{n P_{3/2}\|\mathbf{d}\|n S_{1/2}}\rangle|^2}{|\langle{n P_{1/2}\|\mathbf{d}\|n S_{1/2}}\rangle|^2}.
\end{equation}
With the help of the formula \cite{tensor books}
\begin{multline}\label{n45d}
\langle nJ\|\mathbf{d}\|n'J'\rangle
=(-1)^{L+S+J'+1}\sqrt{(2J+1)(2J'+1)}\\
\times\bigg\{\begin{array}{ccc}
J &1 &J' \\
L' & S& L
\end{array}\bigg\}
\langle nL\|\mathbf{d}\|n'L'\rangle,
\end{multline}
we find
\begin{equation}\label{n45e}
\frac{|\langle{n P_{3/2}\|\mathbf{d}\|n S_{1/2}}\rangle|^2}{|\langle{n P_{1/2}\|\mathbf{d}\|n S_{1/2}}\rangle|^2}=2.
\end{equation}
This explains why the relation $\Delta_2\simeq-2\Delta_1$ is observed for the position of $\lambda_v$ in the case of Fig. \ref{fig6}. 
The deviation of the ratio $\Delta_2/\Delta_1$ from the value of $-2$ is due to several reasons.
The first reason is that a large number of transitions are included in our numerical calculations.
The second reason is that the counter-rotating terms are taken into account in our calculations.
The third reason is that we used the experimental values $|\langle{6 P_{3/2}\|\mathbf{d}\|6 S_{1/2}}\rangle|=6.324$ a.u. and 
$|\langle{6 P_{1/2}\|\mathbf{d}\|6 S_{1/2}}\rangle|=4.489$ a.u., with the ratio
$|\langle{6 P_{3/2}\|\mathbf{d}\|6 S_{1/2}}\rangle|^2/|\langle{6 P_{1/2}\|\mathbf{d}\|6 S_{1/2}}\rangle|^2=1.98$.
The deviation of this ratio from the value of 2 is due to relativistic effects \cite{Sieradzan}.

\section{Summary}
\label{sec:summary}

We provided a concise, yet comprehensive compilation of the general theoretical framework required for calculating the polarizability of the states of multilevel atoms in light fields with arbitrary polarization. Special emphasis is placed on the interpretation of the vector light shift as the result of the action of a fictitious magnetic field. We exemplarily applied the presented formalism to atomic cesium and calculated the scalar, vector, and tensor polarizabilities of the states associated with the D2-line transition. Using these results, we highlighted points of experimental interest such as the red- and blue-detuned magic wavelengths and a wavelength at which the scalar light shift of the ground state vanishes while the vector light shift is substantial. The underlying set of atomic data for the calculations of the polarizability of cesium is explicitly given in tabular as well as electronic forms. By providing all general tools and definitions in a single source and by discussing their respective range of validity, our work should facilitate the theoretical modeling of the light-induced potentials experienced by atoms in complex far-off-resonance optical fields, encountered, e.g., in nonparaxial or near-field optical dipole traps.


\begin{acknowledgments}
We thank B. Arora, R. Grimm, and H. J. Kimble for helpful discussions. 
We are indebted to B. Arora and B. K. Sahoo for giving us the resonance wavelengths and reduced matrix elements for the transitions from the states $6S_{1/2}$ and $6P_{3/2}$ to the states with high principal quantum numbers in atomic cesium.
Financial support by the Wolfgang Pauli Institute is gratefully acknowledged.
\end{acknowledgments}

\appendix

\section{ac Stark shift of a two-level atom interacting with a far-off-resonance light field}
\label{sec:two-level atom}

We consider a two-level atom interacting with a far-off-resonance light field.
Let $|a\rangle$ and $|b\rangle$ be the bare eigenstates of the atom, with unperturbed energies $E_a=\hbar\omega_a$ and $E_b=\hbar\omega_b$, respectively, and let
$\omega$, $\mathcal{E}$, and $\mathbf{u}$ be the frequency, the complex amplitude, and the complex polarization vector, respectively, of the light field. 
The electric component of the light field is given by Eq. (\ref{n1}).
The interaction between the atom and the field is given, in the dipole approximation, by Eq. (\ref{m1}).
The evolution of the off-diagonal density-matrix element $\rho_{ba}$ of the atom is
governed by the equation 
\begin{eqnarray}
\dot{\rho}_{ba}&=&-i(\omega_{b}-\omega_a-i\gamma_{ba}/2)\rho_{ba}
-\frac{i}{2\hbar}(\mathcal{E}\mathbf{u}e^{-i\omega t}
\nonumber\\&&\mbox{}
+\mathcal{E}^*\mathbf{u}^*e^{i\omega t})\cdot\mathbf{d}_{ba}(\rho_{bb}-\rho_{aa}),  
 \label{nano39}
\end{eqnarray}
where $\mathbf{d}_{ba}=\langle b|\mathbf{d}|a\rangle$ is the matrix element of the electric dipole operator 
$\mathbf{d}=\mathbf{d}_{ba}|b\rangle\langle a|+\mathbf{d}_{ab}|a\rangle\langle b|$ 
and $\gamma_{ba}$ is the linewidth of the atomic transition $|b\rangle\leftrightarrow|a\rangle$. In general, we have $\gamma_{ba}=\gamma_{b}+\gamma_{a}$,
where $\gamma_{b}$ and $\gamma_{a}$ are the decay rates of the populations of the levels $|b\rangle$ and $|a\rangle$, respectively. 

We assume that the atom is initially in the level $|a\rangle$, which can be, in general, higher or lower than the level $|b\rangle$.
When the magnitude of the detuning $\omega-|\omega_{b}-\omega_a|$ is large compared to the atomic decay rate $\gamma_{ba}$ and to the magnitude of the Rabi frequency 
$\Omega=d_{ba}\mathcal{E}/\hbar$, we have $\rho_{bb}\simeq 0$ and $\rho_{aa}\simeq 1$. 
We use the ansatz $\rho_{ba}=\rho_{ba}^+e^{-i\omega t}+\rho_{ba}^-e^{i\omega t}$ and assume that $\rho_{ba}^+$ and $\rho_{ba}^-$ vary slowly in time. Then, we find
\begin{eqnarray}
\rho_{ba}^+&=&\frac{\mathcal{E}\mathbf{u}\cdot\mathbf{d}_{ba}}{2\hbar}\frac{1}{\omega_{ba}-\omega-i\gamma_{ba}/2},\nonumber\\
\rho_{ba}^-&=&\frac{\mathcal{E}^*\mathbf{u}^*\cdot\mathbf{d}_{ba}}{2\hbar}\frac{1}{\omega_{ba}+\omega-i\gamma_{ba}/2}.
\label{nano40}
\end{eqnarray}

The induced dipole is given by $\mathbf{p}\equiv\langle\mathbf{d}\rangle=\mathbf{d}_{ba}\rho_{ab}+\mathbf{d}_{ab}\rho_{ba}$. 
It can be written in the form $\mathbf{p}=(\boldsymbol{\wp}e^{-i\omega t}+\boldsymbol{\wp}^*e^{i\omega t})/2$,
where $\boldsymbol{\wp}=2(\mathbf{d}_{ba}\rho_{ba}^{-*}+\mathbf{d}_{ab}\rho_{ba}^+ )$ is the envelope of the positive frequency component. 
We find
\begin{eqnarray}
\boldsymbol{\wp}&=&
\mathbf{d}_{ab}\frac{\mathcal{E}\mathbf{u}\cdot\mathbf{d}_{ba}}{\hbar}\frac{1}{\omega_{ba}-\omega-i\gamma_{ba}/2}
\nonumber\\&&\mbox{}
+\mathbf{d}_{ba}\frac{\mathcal{E}\mathbf{u}\cdot\mathbf{d}_{ab}}{\hbar}\frac{1}{\omega_{ba}+\omega+i\gamma_{ba}/2}.
\label{nano41}
\end{eqnarray}

The ac Stark shift $\delta E_a$ of the energy level $|a\rangle$ is
the time-averaged potential of the induced dipole moment $\mathbf{p}$ interacting with the 
driving electric field $\mathbf{E}$ and is given by
\begin{equation}
\delta E_a=-\frac{1}{2}\overline{\mathbf{p}(t)\cdot \mathbf{E}(t)} =-\frac{1}{4} \mathrm{Re} [\boldsymbol{\wp}\cdot\mathcal{E}^*\mathbf{u}^*].
\label{nano34}
\end{equation}
Here, the factor of $1/2$ accounts for the fact that the dipole moment is induced. Inserting Eq. (\ref{nano41}) into Eq. (\ref{nano34}) yields
\begin{equation}
\delta E_a=  -\frac{|\mathcal{E}|^2}{4\hbar}\mathrm{Re} \bigg(
\frac{|\mathbf{u}\cdot\mathbf{d}_{ba}|^2}{\omega_{ba}-\omega-i\gamma_{ba}/2} 
+\frac{|\mathbf{u}\cdot\mathbf{d}_{ab}|^2}{\omega_{ba}+\omega+i\gamma_{ba}/2}\bigg).
\label{n48}
\end{equation}
We emphasize that Eq. (\ref{n48}) is valid for an arbitrary polarization of the light field.
When we generalize Eq.~(\ref{n48}) to the case of a multilevel atom, we obtain Eq.~(\ref{m2}).

\section{ac Stark interaction operator and components of the dynamical polarizability}
\label{sec:derivation}

In this Appendix, we present the details of the derivation of the expressions for the ac Stark interaction operator $V^{EE}$ and the dynamical scalar, vector, and tensor polarizabilities [see Eqs. (\ref{n10})--(\ref{n6})].
For this purpose, we follow closely Ref. \cite{Rosenbusch09}. 

We use the Cartesian coordinate frame $\{x,y,z\}$.
We introduce the notations
\begin{eqnarray}\label{n49a1} 
A_{-1}&=&(A_{x}-iA_{y})/\sqrt{2}, \nonumber\\ 
A_0&=& A_{z}, \nonumber\\ 
A_{1}&=& -(A_{x}+iA_{y})/\sqrt{2}
\end{eqnarray}
for the spherical tensor components of an arbitrary complex vector $\mathbf{A}=\{A_x,A_y,A_z\}$. In terms of the tensor components $A_q\equiv A_{1q}$, with $q=-1,0,1$, the vector $\mathbf{A}\equiv \mathbf{A}_1$ is an irreducible tensor of rank 1. We introduce the notation $\{\mathbf{A}\otimes\mathbf{B}\}_{K}$ for the irreducible tensor products of rank $K=0,1,2$ 
of two arbitrary vectors $\mathbf{A}$ and $\mathbf{B}$. The $q$ component of the tensor product $\{\mathbf{A}\otimes\mathbf{B}\}_{K}$ is defined as
\begin{equation}\label{n49a}
\{\mathbf{A}\otimes\mathbf{B}\}_{Kq}=\sum_{q_1q_2} C_{1q_11q_2}^{Kq}A_{q_1}B_{q_2},
\end{equation} 
where 
\begin{equation}\label{n49a2}
C_{j_1m_1j_2m_2}^{jm}=(-1)^{j_1-j_2+m}\sqrt{2j+1}\bigg(\begin{array}{ccc}j_1&j_2&j\\m_1&m_2&-m\end{array}\bigg) 
\end{equation} 
is the notation for the Clebsch-Gordan coefficients.
More general, an irreducible tensor product of two irreducible tensors $U_{K_1}$ and $V_{K_2}$ is defined as the irreducible tensor $\{U_{K_1}\otimes V_{K_2}\}_K$ of rank $K$ whose components can be expressed in terms of $U_{K_1q_1}$ and  $V_{K_2q_2}$ according to 
\begin{equation}\label{n49e}
\{U_{K_1}\otimes V_{K_2}\}_{Kq}=\sum_{q_1q_2} C_{K_1q_1K_2q_2}^{Kq}U_{K_1q_1}V_{K_2q_2},
\end{equation}
with $K=|K_1-K_2|,|K_1-K_2|+1,\dots,K_1+K_2-1,K_1+K_2$ and $q=-K,-K+1,\dots,K-1,K$. 
Meanwhile, the scalar product of two irreducible tensors $U_K$ and $V_K$ is defined as 
\begin{equation}\label{n49b}
(U_K\cdot V_K)=\sum_q (-1)^q U_{K,q} V_{K,-q}. 
\end{equation}
 
When we use the formula \cite{tensor books}
\begin{equation}\label{n49d}
(\mathbf{A}\cdot\mathbf{B})(\mathbf{A}'\cdot\mathbf{B}')=\sum_{K=0,1,2} (-1)^K \{\mathbf{A}\otimes\mathbf{A}'\}_K\cdot\{\mathbf{B}\otimes\mathbf{B}'\}_K,
\end{equation} 
which is valid for commuting vectors, we can change the order of coupling of the operators in Eq. (\ref{m3}) to obtain
\begin{eqnarray}\label{n49}
V^{EE}&=& \frac{|\mathcal{E}|^2}{4}\sum_{K=0,1,2}(-1)^K\{\mathbf{u}^*\otimes\mathbf{u}\}_K\cdot[ \{\mathbf{d}\otimes \mathcal{R}_+\mathbf{d}\}_K
\nonumber\\&&\mbox{}
+(-1)^K\{\mathbf{d}\otimes \mathcal{R}_-\mathbf{d}\}_K].
\end{eqnarray}
In deriving the above equation we have employed $\{\mathbf{u}\otimes\mathbf{u}^*\}_K=(-1)^K\{\mathbf{u}^*\otimes\mathbf{u}\}_K$.
When we use the definition (\ref{n49b}) for the scalar product of tensors, we can rewrite Eq. (\ref{n49}) as
\begin{multline}\label{n49c}
V^{EE}= \frac{|\mathcal{E}|^2}{4}\sum_{K=0,1,2}(-1)^K \sum_q (-1)^q\{\mathbf{u}^*\otimes\mathbf{u}\}_{Kq}\\
\times[\{\mathbf{d}\otimes \mathcal{R}_+\mathbf{d}\}_{K,-q}+(-1)^K\{\mathbf{d}\otimes \mathcal{R}_-\mathbf{d}\}_{K,-q}].
\end{multline}
The explicit expressions for the compound tensor components $\{\mathbf{u}^*\otimes\mathbf{u}\}_{Kq}$, which appear in Eqs. (\ref{n49}) and (\ref{n49c}), are
\begin{equation}
\{\mathbf{u}^*\otimes\mathbf{u}\}_{0,0}=-\frac{1}{\sqrt3},
\label{n13}
\end{equation}
\begin{eqnarray}
\{\mathbf{u}^*\otimes\mathbf{u}\}_{1,0}=\frac{|u_1|^2-|u_{-1}|^2}{\sqrt2},
\nonumber\\
\{\mathbf{u}^*\otimes\mathbf{u}\}_{1,1}=-\frac{u_0u^*_{-1}+u_0^*u_1}{\sqrt2},
\nonumber\\
\{\mathbf{u}^*\otimes\mathbf{u}\}_{1,-1}=\frac{u_0u^*_1+u_0^*u_{-1}}{\sqrt2},
\label{n14}
\end{eqnarray}
and
\begin{eqnarray}
\{\mathbf{u}^*\otimes\mathbf{u}\}_{2,0}=\frac{3|u_{0}|^2-1}{\sqrt6},
\nonumber\\
\{\mathbf{u}^*\otimes\mathbf{u}\}_{2,1}=-\frac{u_0u^*_{-1}-u^*_{0}u_1}{\sqrt2},
\nonumber\\
\{\mathbf{u}^*\otimes\mathbf{u}\}_{2,-1}=-\frac{u_0u^*_{1}-u^*_{0}u_{-1}
}{\sqrt2},
\nonumber\\
\{\mathbf{u}^*\otimes\mathbf{u}\}_{2,2}=-u_{1}u^*_{-1},
\nonumber\\
\{\mathbf{u}^*\otimes\mathbf{u}\}_{2,-2}=-u_{-1}u^*_{1}.
\label{n15}
\end{eqnarray}
The operators $\mathcal{R}_+$ and $\mathcal{R}_-$ in Eqs. (\ref{n49}) and (\ref{n49c}) are given by Eqs. (\ref{m4}).
In our treatment given below, the basis states $|a\rangle$ and $|b\rangle$ in Eqs. (\ref{m4})
are taken from the $F$ basis states $|nJFM\rangle$, with unperturbed energies $\omega_{nJFM}=\omega_{nJ}$ and spontaneous decay rates $\gamma_{nJFM}=\gamma_{nJ}$.

Let $V^{EE}_{FMF'M'}\equiv\langle (nJ)FM|V^{EE}| (nJ)F'M'\rangle$ be the matrix elements 
of the Stark interaction operator $V^{EE}$ in the atomic hfs basis $\{|(nJ)FM\rangle\}$ for a fixed set of quantum numbers $nJ$.
From Eq. (\ref{n49c}), we find
\begin{eqnarray}\label{n55}
V^{EE}_{FMF'M'}&=&\frac{|\mathcal{E}|^2}{4}\sum_{K=0,1,2}(-1)^K \sum_q (-1)^q\{\mathbf{u}^*\otimes\mathbf{u}\}_{Kq}
\nonumber\\&&\mbox{}
\times \mathcal{O}_{FMF'M'}^{Kq},
\end{eqnarray}
where
\begin{eqnarray}\label{n56}
\lefteqn{\mathcal{O}_{FMF'M'}^{Kq}=\sum_{q_1q_2}C^{K,-q}_{1q_11q_2}} 
\nonumber\\&&\mbox{}\times
\sum_{n''J''F''M''}
\langle nJFM| d_{q_1} |n''J''F''M''\rangle 
\nonumber\\&&\mbox{}\times
\langle n''J''F''M''| d_{q_2}|nJF'M'\rangle
\mathcal{R}^{(K)}_{n''J''nJ},
\end{eqnarray}
with
\begin{eqnarray}\label{n57}
\mathcal{R}^{(K)}_{n''J''nJ}&=&
-\frac{1}{\hbar}\mathrm{Re}\bigg(\frac{1}{\omega_{n''J''}-\omega_{nJ}-\omega-i\gamma_{n''J''nJ}/2}
\nonumber\\&&\mbox{}
+\frac{(-1)^K}{\omega_{n''J''}-\omega_{nJ}+\omega+i\gamma_{n''J''nJ}/2}\bigg).\qquad
\end{eqnarray}

According to the Wigner-Eckart theorem \cite{tensor books}, the dependence of the matrix elements $\langle nJFM|T_{Kq}| n'J'F'M'\rangle$ of tensor component operators $T_{Kq}$ on the quantum numbers $M$, $M'$, and $q$ is entirely included in the Wigner  3-$j$ symbol, namely,
\begin{multline}\label{n50}
\langle nJFM|T_{Kq}|n'J'F'M'\rangle=\\ 
(-1)^{F-M}
\bigg(\begin{array}{ccc}
F &K &F' \\
-M & q& M'
\end{array}\bigg)
\langle nJF \| T_K\|n'J'F' \rangle.
\end{multline}
Here, the invariant factor 
\begin{multline}\label{n50a}
\langle nJF \| T_K\|n'J'F' \rangle
=\sum_{MM'q}(-1)^{F-M}
\bigg(\begin{array}{ccc}
F &K &F' \\
-M & q& M'
\end{array}\bigg)\\
\times\langle nJFM|T_{Kq}|n'J'F'M'\rangle
\end{multline}
is the reduced matrix element for the set of tensor component operators $T_{Kq}$,
with the normalization convention
\begin{equation}\label{n50b}
|\langle nJF \| T_K\|n'J'F' \rangle|^2
=\sum_{MM'q}
|\langle nJFM|T_{Kq}|n'J'F'M'\rangle|^2
\end{equation}
and the complex conjugate relation
\begin{equation}\label{n50c}
\langle nJF \| T_K\|n'J'F' \rangle^*
=(-1)^{F-F'}\langle n'J'F' \| T_K\|nJF \rangle.
\end{equation}
Since the electric dipole $\mathbf{d}$ is a tensor of rank 1, the application of the Wigner-Eckart theorem to the matrix elements $\langle nJFM|d_q| n'J'F'M'\rangle$ of the spherical-tensor-component operators $d_q$ of the electric dipole gives
\begin{eqnarray}\label{n58}
\lefteqn{\langle nJFM|d_q|n'J'F'M'\rangle}
\nonumber\\&&
=(-1)^{F-M}
\bigg(\begin{array}{ccc}
F &1 &F' \\
-M & q& M'
\end{array}\bigg)
\langle nJF \| \mathbf{d}\|n'J'F' \rangle.\quad
\end{eqnarray}
The invariant factor $\langle nJF \| \mathbf{d}\|n'J'F' \rangle$ is the reduced matrix element for the electric dipole operator $\mathbf{d}$. 
With the help of Eq. (\ref{n58}), we can rewrite Eq. (\ref{n56}) as 
\begin{eqnarray}\label{n59}
\lefteqn{\mathcal{O}_{FMF'M'}^{Kq}=\sum_{n''J''F''}
\langle nJF\| \mathbf{d}\|n''J''F''\rangle} 
\nonumber\\&&\mbox{}\times
\langle n''J''F''\| \mathbf{d}\|nJF'\rangle
\mathcal{R}^{(K)}_{n''J''nJ}\mathcal{N}^{KqF''}_{FMF'M'},
\end{eqnarray}
where
\begin{eqnarray}\label{n60}
\mathcal{N}^{KqF''}_{FMF'M'}&=&\sqrt{2K+1}
\sum_{q_1q_2M''}(-1)^{F+F''-M-M''-q}
\nonumber\\&&\mbox{}\times 
\bigg(\begin{array}{ccc}
1   &1  &K \\
q_1 &q_2 & q
\end{array}\bigg)
\bigg(\begin{array}{ccc}
F  &1  &F'' \\
-M &q_1 &M''
\end{array}\bigg)
\nonumber\\&&\mbox{}\times 
\bigg(\begin{array}{ccc}
F'' &1   & F' \\
-M''&q_2 & M' 
\end{array}\bigg).
\end{eqnarray}
When we use the symmetry properties of the 3-$j$ symbol and  
the sum rule \cite{tensor books}
\begin{multline}\label{n61}
\sum_{m_4m_5m_6}(-1)^{j_4+j_5+j_6-m_4-m_5-m_6}
\bigg(\begin{array}{ccc}
j_5  &j_1  &j_6 \\
m_5 &-m_1 &-m_6
\end{array}\bigg)\\ 
\times\bigg(\begin{array}{ccc}
j_6  &j_2  &j_4 \\
m_6 &-m_2 &-m_4
\end{array}\bigg)
\bigg(\begin{array}{ccc}
j_4  &j_3  &j_5 \\
m_4 &-m_3 &-m_5
\end{array}\bigg)\\ 
= \bigg(\begin{array}{ccc}
j_1  &j_2  &j_3 \\
m_1 &m_2 &m_3
\end{array}\bigg)
\bigg\{\begin{array}{ccc}
j_1  &j_2  &j_3 \\
j_4 &j_5 &j_6
\end{array}\bigg\},
\end{multline}
we find
\begin{eqnarray}\label{n62}
\mathcal{N}^{KqF''}_{FMF'M'}&=&(-1)^{K+F'+M}\sqrt{2K+1}
\bigg(\begin{array}{ccc}
 F &K  &F' \\
-M &-q & M'
\end{array}\bigg)
\nonumber\\&&\mbox{}\times
\bigg\{\begin{array}{ccc}
1 &K   &1 \\
F &F'' &F'
\end{array}\bigg\}.
\end{eqnarray}
We now insert Eq. (\ref{n62}) into Eq. (\ref{n59}) and then insert the result into Eq. (\ref{n55}). Then, we obtain 
\cite{Rosenbusch09} 
\begin{multline}\label{n51}
V^{EE}_{FMF'M'}= \frac{|\mathcal{E}|^2}{4}\sum_{K=0,1,2}(-1)^K \sum_q (-1)^q\{\mathbf{u}^*\otimes\mathbf{u}\}_{Kq}\\
\times(-1)^{F-M}\bigg(\begin{array}{ccc}
F &K &F' \\
-M & -q& M'
\end{array}\bigg)
\alpha^{(K)}_{nJFF'},
\end{multline}
where
\begin{multline}\label{n64}
\alpha^{(K)}_{nJFF'}=(-1)^{K+F+F'}
\sqrt{2K+1}\sum_{n''J''F''} 
\bigg\{\begin{array}{ccc}
1 &K   &1 \\
F &F'' &F'
\end{array}\bigg\}\\ 
\times\langle nJF\| \mathbf{d}\|n''J''F''\rangle 
\langle n''J''F''\| \mathbf{d}\|nJF'\rangle
\mathcal{R}^{(K)}_{n''J''nJ}
\end{multline}
are the reduced scalar ($K=0$), vector ($K=1$), and tensor ($K=2$) polarizability coefficients for the  hfs levels within a fine-structure manifold $nJ$.

For the tensor component operators $T_{Kq}$ that do not act on the nuclear spin degrees of freedom, the dependence of the reduced matrix element $\langle nJIF\|T_K\|n'J'I'F'\rangle$ on $F$, $F'$, $I$, and $I'$ may be factored out as \cite{tensor books}
\begin{multline}\label{n53a}
\langle nJIF\|T_K\|n'J'I'F'\rangle=\delta_{II'}(-1)^{J+I+F'+K}\\
\times\sqrt{(2F+1)(2F'+1)}\bigg\{\begin{array}{ccc}
F &K &F' \\
J' & I& J
\end{array}\bigg\}
\langle nJ\|T_K\|n'J'\rangle.
\end{multline}
Since the electric dipole $\mathbf{d}$ of the atom does not couple to the nuclear degrees of freedom and is a tensor of rank 1, the use of Eq. (\ref{n53a}) for the case $T_K=\mathbf{d}$ yields \cite{tensor books}
\begin{multline}\label{n65}
\langle nJF\|\mathbf{d}\|n'J'F'\rangle\equiv\langle nJIF\|\mathbf{d}\|n'J'IF'\rangle\\
=(-1)^{J+I+F'+1}\sqrt{(2F+1)(2F'+1)}\\
\times\bigg\{\begin{array}{ccc}
F &1 &F' \\
J' & I& J
\end{array}\bigg\}
\langle nJ\|\mathbf{d}\|n'J'\rangle.
\end{multline}
Substituting Eq. (\ref{n65}) into Eq. (\ref{n64}) yields 
\begin{multline}\label{n67}
\alpha^{(K)}_{nJFF'}=(-1)^{K+J+2I+F+2F'}
\sqrt{(2F+1)(2F'+1)}
\\\times
\sum_{n''J''} (-1)^{J''}
\langle nJ\|\mathbf{d}\|n''J''\rangle
\langle n''J''\|\mathbf{d}\|nJ\rangle
\mathcal{R}^{(K)}_{n''J''nJ}
\\\times
\sqrt{2K+1}\sum_{F''} (-1)^{F''}(2F''+1)
\bigg\{\begin{array}{ccc}
1 &F' &F'' \\
F &1  &K
\end{array}\bigg\}\\ 
\times\bigg\{\begin{array}{ccc}
F   &1 &F'' \\
J'' &I &J
\end{array}\bigg\}
\bigg\{\begin{array}{ccc}
J'' &I &F'' \\
F'  &1  &J  
\end{array}\bigg\}.
\end{multline}
The summation over $F''$ in Eq. (\ref{n67}) can be performed using the formula \cite{tensor books}
\begin{multline}\label{n66}
\sum_{k}(-1)^{k}(2k+1)
\bigg\{\begin{array}{ccc}
j_1 &j_2 &k \\
j_3 &j_4 &j_5
\end{array}\bigg\}
\bigg\{\begin{array}{ccc}
j_3 &j_4 &k \\
j_6 &j_7 &j_8
\end{array}\bigg\}
\\ \times
\bigg\{\begin{array}{ccc}
j_6 &j_7 &k \\
j_2 &j_1 &j_9
\end{array}\bigg\} 
=(-1)^{-j_1-j_2-j_3-j_4-j_5-j_6-j_7-j_8-j_9}\\ 
\times\bigg\{\begin{array}{ccc}
j_5 &j_8 &j_9 \\
j_6 &j_1 &j_4
\end{array}\bigg\}
\bigg\{\begin{array}{ccc}
j_5 &j_8 &j_9 \\
j_7 &j_2 &j_3
\end{array}\bigg\}.
\end{multline}
The result is
\begin{multline}\label{n68}
\alpha^{(K)}_{nJFF'}=(-1)^{I+F'-J}
\sqrt{(2F+1)(2F'+1)}
\\\times
\sum_{n''J''} 
\langle nJ\|\mathbf{d}\|n''J''\rangle
\langle n''J''\|\mathbf{d}\|nJ\rangle
\mathcal{R}^{(K)}_{n''J''nJ}
\\\times
\sqrt{2K+1}
\bigg\{\begin{array}{ccc}
1 &K   &1 \\
J &J'' &J
\end{array}\bigg\}
\bigg\{\begin{array}{ccc}
F &K &F'\\
J &I &J 
\end{array}\bigg\}.
\end{multline}
When we insert the explicit expression (\ref{n57}) into the above equation, we get \cite{Rosenbusch09}
\begin{equation}\label{n53b}
\begin{split}
\alpha^{(K)}_{nJFF'}=&(-1)^{J+I+F'+K}\sqrt{(2F+1)(2F'+1)}\\
&\times\bigg\{\begin{array}{ccc}
F &K &F' \\
J & I& J
\end{array}\bigg\}
\alpha^{(K)}_{nJ},
\end{split}
\end{equation}
where 
\begin{equation}\label{n54}
\begin{split}
\alpha^{(K)}_{nJ}=&(-1)^{2J+K+1}\sqrt{2K+1}
\sum_{n'J'} 
\bigg\{\begin{array}{ccc}
1 &K &1 \\
J & J'& J
\end{array}\bigg\}\\
&\times
\langle nJ\|\mathbf{d}\|n'J'\rangle\langle n'J'\|\mathbf{d}\|nJ\rangle\\
&\times
\frac{1}{\hbar}\mathrm{Re}\bigg(\frac{1}{\omega_{n'J'}-\omega_{nJ}-\omega-i\gamma_{n'J'nJ}/2}\\
&+\frac{(-1)^K}{\omega_{n'J'}-\omega_{nJ}+\omega+i\gamma_{n'J'nJ}/2}\bigg)
\end{split}
\end{equation}
are the reduced scalar ($K=0$), vector ($K=1$), and tensor ($K=2$) polarizabilities for the Stark shift of  the fine-structure level $|nJ\rangle$. When we substitute Eq. (\ref{n53b}) into Eq. (\ref{n51}), we obtain Eq. (\ref{n11}).
Since $\langle nJ\|\mathbf{d}\|n'J'\rangle=(-1)^{J-J'}\langle n'J'\|\mathbf{d}\|nJ\rangle$,
Eq. (\ref{n54}) can be rewritten as Eq. (\ref{n6}). 
The explicit expressions for the reduced polarizabilities $\alpha^{(K)}_{nJ}$ are
\begin{eqnarray}
\alpha^{(0)}_{nJ}
&=&\frac{2}{\hbar\sqrt{3(2J+1)}}\sum_{n'J'}
|\langle{n' J'\|\mathbf{d}\|n J}\rangle|^2
\nonumber\\&&\mbox{}\times
\frac{\omega_{n'J'nJ}(\omega_{n'J'nJ}^2
-\omega^2+\gamma_{n'J'nJ}^2/4)}{(\omega_{n'J'nJ}^2
-\omega^2+\gamma_{n'J'nJ}^2/4)^2+\gamma_{n'J'nJ}^2\omega^2},
\nonumber\\
\alpha^{(1)}_{nJ}
&=&\frac{2\sqrt{3}}{\hbar}\sum_{n'J'}(-1)^{J+J'}
\bigg\{\begin{array}{ccc}
1 &1 &1 \\
J &J'&J 
\end{array}\bigg\}
|\langle{n' J'\|\mathbf{d}\|n J}\rangle|^2
\nonumber\\&&\mbox{}\times
\frac{\omega(\omega_{n'J'nJ}^2
-\omega^2-\gamma_{n'J'nJ}^2/4)}{(\omega_{n'J'nJ}^2
-\omega^2+\gamma_{n'J'nJ}^2/4)^2+\gamma_{n'J'nJ}^2\omega^2},
\nonumber\\
\alpha^{(2)}_{nJ}
&=&-\frac{2\sqrt{5}}{\hbar}
\sum_{n'J'}(-1)^{J+J'}
\bigg\{\begin{array}{ccc}
1 &2 &1 \\
J &J'&J 
\end{array}\bigg\}
|\langle{n' J'\|\mathbf{d}\|n J}\rangle|^2
\nonumber\\&&\mbox{}\times
\frac{\omega_{n'J'nJ}(\omega_{n'J'nJ}^2
-\omega^2+\gamma_{n'J'nJ}^2/4)}{(\omega_{n'J'nJ}^2
-\omega^2+\gamma_{n'J'nJ}^2/4)^2+\gamma_{n'J'nJ}^2\omega^2}.
\nonumber\\
\label{n7}
\end{eqnarray}
When we neglect the linewidths $\gamma_{n'J'nJ}$, Eqs. (\ref{n54}) and (\ref{n7}) come to full agreement with the results of Ref. \cite{Rosenbusch09}.

We emphasize that, in the above calculations for the polarizabilities, we used the approximation $\omega_{nJFM}=\omega_{nJ}$, that is, we neglected the effect of the hfs splitting on the polarizabilities. This approximation is consistent with the perturbation theory scheme used in our case where the hfs splitting and the ac Stark shift are considered to be small perturbations of the same order. If the hfs splitting is much larger than the ac Stark shift, we can consider only the ac Stark shift as a perturbation.  
In this case, the operator for the ac Stark shifts of sublevels of a hfs level $|nJF\rangle$ is given by expression \cite{Rosenbusch09}
\begin{eqnarray}
\lefteqn{V^{EE}= -\frac{1}{4}|\mathcal{E}|^2 \bigg\{\alpha^s_F 
-i\alpha^{v}_F \frac{[\mathbf{u}^*\times\mathbf{u}]\cdot\mathbf{F}}{2F}}
\nonumber\\&&\mbox{}
+\alpha^{T}_F\frac{3[(\mathbf{u}^*\cdot\mathbf{F})(\mathbf{u}\cdot\mathbf{F})
+(\mathbf{u}\cdot\mathbf{F})(\mathbf{u}^*\cdot\mathbf{F})]
-2\mathbf{F}^2}{2F(2F-1)}\bigg\},\nonumber\\
\label{n71}
\end{eqnarray}
with the matrix elements
\begin{multline}\label{n72}
V^{EE}_{MM'}= \frac{|\mathcal{E}|^2}{4}\sum_{K=0,1,2}(-1)^K \sum_q (-1)^q\{\mathbf{u}^*\otimes\mathbf{u}\}_{Kq}\\
\times(-1)^{F-M}\bigg(\begin{array}{ccc}
F &K &F \\
-M & -q& M'
\end{array}\bigg)
\alpha^{(K)}_{F},
\end{multline}
where
\begin{eqnarray}
\alpha^s_{F}&=&\frac{1}{\sqrt{3(2F+1)}} \alpha^{(0)}_{F},
\nonumber\\
\alpha^v_{F}&=&-\sqrt{\frac{2F}{(F+1)(2F+1)}}\alpha^{(1)}_{F},
\nonumber\\
\alpha^T_{F}&=&-\sqrt{\frac{2F(2F-1)}{3(F+1)(2F+1)(2F+3)}}\alpha^{(2)}_{F},
\label{n72a}
\end{eqnarray}
and
\begin{multline}\label{n73}
\alpha^{(K)}_{F}=(-1)^{K+F+1}
(2F+1)\sqrt{2K+1}
\sum_{n'J'}
|\langle n'J'\|\mathbf{d}\|nJ\rangle|^2
\\\times
\sum_{F'} (-1)^{F'}(2F'+1)
\bigg\{\begin{array}{ccc}
1 &K  &1\\
F &F' &F 
\end{array}\bigg\}
\bigg\{\begin{array}{ccc}
F  &1 &F' \\
J' &I &J
\end{array}\bigg\}^2\\
\times\frac{1}{\hbar}\mathrm{Re}\bigg(\frac{1}{\omega_{n'J'F'}-\omega_{nJF}-\omega-i\gamma_{n'J'F'nJF}/2}\\
+\frac{(-1)^K}{\omega_{n'J'F'}-\omega_{nJF}+\omega+i\gamma_{n'J'F'nJF}/2}\bigg).
\end{multline}
We note that, in the framework of the validity of Eqs. (\ref{n71}) and (\ref{n73}), the different hfs levels $F=I+1/2$ and $F=I-1/2$ of the ground state have different scalar polarizabilities.
This means that, when the hfs splitting is taken into account in the expression for the atomic transition frequency $\omega_{n'J'F'}-\omega_{nJF}$, a nonzero differential scalar Stark shift between the hfs levels of the ground state may occur. 

\section{Additional magnetic field}
\label{sec:magnetic}

We consider the presence of a weak external real magnetic field $\mathbf{B}$.
The Hamiltonian for the interaction between the magnetic field and the atom is \cite{Metcalf99}
\begin{equation}
V^B=\mu_Bg_{nJ} (\mathbf{J}\cdot\mathbf{B}).
\label{n42}
\end{equation}
It can be shown that the matrix elements of the operator $V^B$ in the basis $\{|FM\rangle\}$ are given by the expression
\begin{eqnarray}
V_{FMF'M'}^{B}&=&\mu_Bg_{nJ}(-1)^{J+I-M}
\sqrt{J(J+1)(2J+1)}
\nonumber\\&&\mbox{}
\times\sqrt{(2F+1)(2F'+1)}
\bigg\{\begin{array}{ccc}
F &1 &F' \\
J &I &J 
\end{array}\bigg\}
\nonumber\\&&\mbox{}
\times
\sum_{{q=0,\pm1}} 
(-1)^{q}B_{q}
\bigg(\begin{array}{ccc}
F &1 &F' \\
M & q& -M'
\end{array}\bigg).
\label{n44}
\end{eqnarray}
Here, $B_{-1}=(B_{x}-iB_{y})/\sqrt{2}$, $B_0= B_{z}$, and $B_{1}= -(B_{x}+iB_{y})/\sqrt{2}$
are the spherical tensor components of the magnetic induction vector $\mathbf{B}=\{B_x,B_y,B_z\}$.
We note that Eq. (\ref{n44}) is valid for an arbitrary quantization axis $z$.

When $F$ is a good quantum number, the interaction operator (\ref{n42}) can be replaced by the operator
\begin{equation}
V^B=\mu_Bg_{nJF} (\mathbf{F}\cdot\mathbf{B}).
\label{n45}
\end{equation}
In the absence of the light field, the energies of the Zeeman sublevels are $\hbar\omega_{nJFM}=\hbar\omega_{nJF}+\mu_B g_{nJF}BM$. 
Here, $\hbar\omega_{nJF}$ is the energy of the hfs level $|nJF\rangle$ in the absence of the magnetic field and $M=-F,\dots,F$ is the magnetic quantum number. This  integer number is an eigenvalue corresponding to the eigenstate $|FM\rangle_B$ of the projection $F_{z_B}$ of $\mathbf{F}$ onto the $z_B$ axis. 
In general, the quantization axis $z$ may be different from the magnetic field axis $z_B$ and, consequently, $|FM\rangle$ may be different from $|FM\rangle_B$. In order to find the level energy shifts, we must add the magnetic interaction operator $V^B$ to the combined hfs-plus-Stark interaction operator (\ref{n40}) and then diagonalize the resulting operator.

\section{Atomic level energies and reduced matrix elements}
\label{sec:data}

The full set of parameters used in our numerical calculations is shown in Tables \ref{data6SP1/2}--\ref{data6PD5/2}. The data is also provided 
as electronic files in \cite{Supplementary}. The reduced matrix elements are given in the atomic unit $ea_0$ of the electric dipole.
The energies of the levels with the principal quantum number $n\le 25$ are taken from \cite{NIST}. 
The energies of the levels with the principal quantum number $n\ge 26$ are provided by Arora and Sahoo \cite{Bindiya}. The reduced matrix elements for the transitions $6S_{1/2}\leftrightarrow (6\text{--}15)P_{1/2,3/2}$ are taken from \cite{Sieradzan}. 
The reduced matrix elements for the transitions $6P_{3/2}\leftrightarrow (6\text{--}10)S_{1/2}$ and $6P_{3/2}\leftrightarrow (5\text{--}8)D_{3/2,5/2}$ are taken from \cite{Arora07}. The reduced matrix elements for transitions to highly excited states are provided by Arora and Sahoo \cite{Bindiya}.


 \begin{table}[H] 
 \caption{\label{data6SP1/2} Energies of levels $nP_{1/2}$ and reduced matrix elements for transitions between levels $nP_{1/2}$ and $6S_{1/2}$ in atomic cesium.
 The data of this table is provided as an electronic file in \cite{Supplementary}.}
 \begin{tabular}{ccc}
 \hline\noalign{\smallskip}
 $nP_{1/2}$ level & Energy             & $|\langle nP_{1/2}\|\mathbf{d}\|6S_{1/2}\rangle|$  \\
                  & (cm$^{-1}$)        & (a.u.) \\
 \hline\noalign{\smallskip}

           6$P_{1/2}$ &   11178.27     &   4.489      \\
           7$P_{1/2}$ &   21765.35     &  0.276     \\
           8$P_{1/2}$ &   25708.84     &  0.081 \\
           9$P_{1/2}$ &   27637.00     &  0.043 \\
          10$P_{1/2}$ &   28726.81     &  0.047 \\
          11$P_{1/2}$ &   29403.42     &  0.034 \\
          12$P_{1/2}$ &   29852.43     &  0.026 \\
          13$P_{1/2}$ &   30165.67     &  0.021 \\
          14$P_{1/2}$ &   30392.87     &  0.017 \\
          15$P_{1/2}$ &   30562.91     &  0.015 \\
          16$P_{1/2}$ &   30693.47     &  0.022 \\
          17$P_{1/2}$ &   30795.91     &  0.023 \\
          18$P_{1/2}$ &   30877.75     &  0.019 \\
          19$P_{1/2}$ &   30944.17     &  0.010 \\
          20$P_{1/2}$ &   30998.79     &  0.035 \\
          21$P_{1/2}$ &   31044.31     &  0.002 \\
          22$P_{1/2}$ &   31082.60     &  0.027 \\
          23$P_{1/2}$ &   31115.12     &  0.002 \\
          24$P_{1/2}$ &   31142.97     &  0.000 \\
          25$P_{1/2}$ &   31167.02     &  0.031 \\
          26$P_{1/2}$ &   79752.85     &  0.045 \\
          27$P_{1/2}$ &   123985.9     &  0.044 \\
          28$P_{1/2}$ &   201030.4     &  0.040 \\
          29$P_{1/2}$ &   334252.0     &  0.035 \\
          30$P_{1/2}$ &   563660.5     &  0.029 \\
          31$P_{1/2}$ &   958103.9     &  0.022 \\
          32$P_{1/2}$ &   1636312     &  0.015 \\
          33$P_{1/2}$ &   2804845     &  0.010 \\
          34$P_{1/2}$ &   4827435     &  0.007 \\
          35$P_{1/2}$ &   8334762     &  0.004 \\
          36$P_{1/2}$ &  14384903 &  0.002 \\
          37$P_{1/2}$ &  19069034 &  0.000 \\
          38$P_{1/2}$ &  24735902 &  0.001 \\
          39$P_{1/2}$ &  42263044 &  0.001 \\
          40$P_{1/2}$ &  71483200 &  0.000 \\
 
\noalign{\smallskip}\hline
\end{tabular}
\end{table}


 \begin{table}[H] 
 \caption{\label{data6SP3/2} Energies of levels $nP_{3/2}$ and reduced matrix elements for transitions between levels $nP_{3/2}$ and $6S_{1/2}$ in atomic cesium. The data of this table is provided as an electronic file in \cite{Supplementary}.}
 \begin{tabular}{ccc}
 \hline\noalign{\smallskip}
 $nP_{3/2}$ level & Energy      & $|\langle nP_{3/2}\|\mathbf{d}\|6S_{1/2}\rangle|$  \\
                  & (cm$^{-1}$) & (a.u.) \\
 \hline\noalign{\smallskip}

           6$P_{3/2}$ &   11732.31     &   6.324      \\
           7$P_{3/2}$ &   21946.40     &  0.586      \\
           8$P_{3/2}$ &   25791.51     &  0.218      \\
           9$P_{3/2}$ &   27681.68     &  0.127      \\
          10$P_{3/2}$ &   28753.68     &  0.114      \\
          11$P_{3/2}$ &   29420.82     &  0.085 \\
          12$P_{3/2}$ &   29864.34     &  0.067 \\
          13$P_{3/2}$ &   30174.18     &  0.055 \\
          14$P_{3/2}$ &   30399.16     &  0.046 \\
          15$P_{3/2}$ &   30567.69     &  0.039 \\
          16$P_{3/2}$ &   30697.19     &  0.062 \\
          17$P_{3/2}$ &   30798.85     &  0.065 \\
          18$P_{3/2}$ &   30880.12     &  0.062 \\
          19$P_{3/2}$ &   30946.11     &  0.039 \\
          20$P_{3/2}$ &   31000.40     &  0.112      \\
          21$P_{3/2}$ &   31045.66     &  0.014 \\
          22$P_{3/2}$ &   31083.77     &  0.119      \\
          23$P_{3/2}$ &   31116.09     &  0.081 \\
          24$P_{3/2}$ &   31143.84     &  0.004 \\
          25$P_{3/2}$ &   31167.74     &  0.032 \\
          26$P_{3/2}$ &   82736.05     &  0.001 \\
          27$P_{3/2}$ &   129247.4     &  0.017 \\
          28$P_{3/2}$ &   210254.6     &  0.025 \\
          29$P_{3/2}$ &   350320.8     &  0.029 \\
          30$P_{3/2}$ &   591569.5     &  0.029 \\
          31$P_{3/2}$ &   1006618     &  0.023 \\
          32$P_{3/2}$ &   1720992     &  0.017 \\
          33$P_{3/2}$ &   2953614     &  0.012 \\
          34$P_{3/2}$ &   5089655     &  0.008 \\
          35$P_{3/2}$ &   8794667     &  0.005 \\
          36$P_{3/2}$ &  15185971 &  0.003 \\
          37$P_{3/2}$ &  19069122 &  0.000 \\
          38$P_{3/2}$ &  26120330 &  0.002 \\
          39$P_{3/2}$ &  44625560 &  0.001 \\
          40$P_{3/2}$ &  75454848 &  0.001 \\
 
\noalign{\smallskip}\hline
\end{tabular}
\end{table}


 \begin{table}[H] 
 \caption{\label{data6PS} Energies of levels $nS_{1/2}$ and reduced matrix elements for transitions between levels $nS_{1/2}$ and $6P_{3/2}$ in atomic cesium. The data of this table is provided as an electronic file in \cite{Supplementary}.}
 \begin{tabular}{ccc}
 \hline\noalign{\smallskip}
 $nS_{1/2}$ level & Energy            & $|\langle nS_{1/2}\|\mathbf{d}\|6P_{3/2}\rangle|$ \\
                  & (cm$^{-1}$)       & (a.u.)\\ 
 \hline\noalign{\smallskip}

           6$S_{1/2}$ &   0            &   6.324      \\
           7$S_{1/2}$ &   18535.53     &   6.470      \\
           8$S_{1/2}$ &   24317.15     &   1.461      \\
           9$S_{1/2}$ &   26910.66     &  0.770      \\
          10$S_{1/2}$ &   28300.23     &  0.509      \\
          11$S_{1/2}$ &   29131.73     &  0.381      \\
          12$S_{1/2}$ &   29668.80     &  0.297      \\
          13$S_{1/2}$ &   30035.79     &  0.241      \\
          14$S_{1/2}$ &   30297.64     &  0.219      \\
          15$S_{1/2}$ &   30491.02     &  0.234      \\
          16$S_{1/2}$ &   30637.88     &  0.251      \\
          17$S_{1/2}$ &   30752.03     &  0.259      \\
          18$S_{1/2}$ &   30842.52     &  0.239      \\
          19$S_{1/2}$ &   30915.45     &  0.376      \\
          20$S_{1/2}$ &   30975.10     &  0.213      \\
          21$S_{1/2}$ &   31024.50     &  0.349      \\
          22$S_{1/2}$ &   31065.88     &  0.472      \\
          23$S_{1/2}$ &   31100.88     &  0.578      \\
          24$S_{1/2}$ &   31130.75     &  0.026      \\
          25$S_{1/2}$ &   31156.44     &  0.464      \\
          26$S_{1/2}$ &   74266.21     &  0.325      \\
          27$S_{1/2}$ &   114033.8     &  0.211      \\
          28$S_{1/2}$ &   184506.6     &  0.132      \\
          29$S_{1/2}$ &   308495.8     &  0.081 \\
          30$S_{1/2}$ &   525668.9     &  0.048 \\
          31$S_{1/2}$ &   905144.4     &  0.027 \\
          32$S_{1/2}$ &   1567994      &  0.015 \\
          33$S_{1/2}$ &   2728852      &  0.008 \\
          34$S_{1/2}$ &   4773681      &  0.004 \\
          35$S_{1/2}$ &   8387611      &  0.002 \\
          36$S_{1/2}$ &  14752595      &  0.001 \\
          37$S_{1/2}$ &  19071256      &  0.000 \\
          38$S_{1/2}$ &  25891848      &  0.001 \\
          39$S_{1/2}$ &  45209868      &  0.000 \\
          40$S_{1/2}$ &  78196928      &  0.000 \\
 
\noalign{\smallskip}\hline
\end{tabular}
\end{table}


 \begin{table}[H] 
 \caption{\label{data6PD3/2} Energies of levels $nD_{3/2}$ and reduced matrix elements for transitions between levels $nD_{3/2}$ and $6P_{3/2}$ in atomic cesium. The data of this table is provided as an electronic file in \cite{Supplementary}.}
 \begin{tabular}{ccc}
 \hline\noalign{\smallskip}
 $nD_{3/2}$ level & Energy            & $|\langle nD_{3/2}\|\mathbf{d}\|6P_{3/2}\rangle|$ \\
                  & (cm$^{-1}$)       & (a.u.)\\ 
 \hline\noalign{\smallskip}

           5$D_{3/2}$ &   14499.26     &   3.166      \\
           6$D_{3/2}$ &   22588.82     &   2.100      \\
           7$D_{3/2}$ &   26047.83     &  0.976      \\
           8$D_{3/2}$ &   27811.24     &  0.607      \\
           9$D_{3/2}$ &   28828.68     &  0.391      \\
          10$D_{3/2}$ &   29468.29     &  0.304      \\
          11$D_{3/2}$ &   29896.34     &  0.246      \\
          12$D_{3/2}$ &   30196.80     &  0.211      \\
          13$D_{3/2}$ &   30415.75     &  0.215      \\
          14$D_{3/2}$ &   30580.23     &  0.234      \\
          15$D_{3/2}$ &   30706.90     &  0.248      \\
          16$D_{3/2}$ &   30806.53     &  0.256      \\
          17$D_{3/2}$ &   30886.30     &  0.269      \\
          18$D_{3/2}$ &   30951.15     &  0.204      \\
          19$D_{3/2}$ &   31004.59     &  0.397      \\
          20$D_{3/2}$ &   31049.14     &  0.012 \\
          21$D_{3/2}$ &   31086.68     &  0.482      \\
          22$D_{3/2}$ &   31118.60     &  0.021 \\
          23$D_{3/2}$ &   31145.97     &  0.491      \\
          24$D_{3/2}$ &   31169.61     &  0.438      \\
          25$D_{3/2}$ &   31190.18     &  0.344      \\
          26$D_{3/2}$ &   53740.33     &  0.028 \\
          27$D_{3/2}$ &   67538.40     &  0.244      \\
          28$D_{3/2}$ &   89286.16     &  0.158      \\
          29$D_{3/2}$ &   122155.2     &  0.095 \\
          30$D_{3/2}$ &   171666.2     &  0.053 \\
          31$D_{3/2}$ &   246044.2     &  0.027 \\
          32$D_{3/2}$ &   357540.4     &  0.011 \\
          33$D_{3/2}$ &   524435.1     &  0.002 \\
          34$D_{3/2}$ &   774031.9     &  0.002 \\
          35$D_{3/2}$ &   1147062      &  0.004 \\
          36$D_{3/2}$ &   1704131      &  0.004 \\
          37$D_{3/2}$ &   2535326      &  0.004 \\
          38$D_{3/2}$ &   3774912      &  0.003 \\
          39$D_{3/2}$ &   5623814      &  0.002 \\
          40$D_{3/2}$ &   8382140      &  0.002 \\
          41$D_{3/2}$ &  12493172      &  0.001 \\
          42$D_{3/2}$ &  18603884      &  0.001 \\
 
\noalign{\smallskip}\hline
\end{tabular}
\end{table}


 \begin{table}[H] 
 \caption{\label{data6PD5/2} Energies of levels $nD_{5/2}$ and reduced matrix elements for transitions between levels $nD_{5/2}$ and $6P_{3/2}$ in atomic cesium. The data of this table is provided as an electronic file in \cite{Supplementary}.}
 \begin{tabular}{ccc}
 \hline\noalign{\smallskip}
 $nD_{5/2}$ level & Energy            & $|\langle nD_{5/2}\|\mathbf{d}\|6P_{3/2}\rangle|$ \\
                  & (cm$^{-1}$)       & (a.u.)\\ 
 \hline\noalign{\smallskip}

           5$D_{5/2}$ &   14596.84     &   9.590      \\
           6$D_{5/2}$ &   22631.69     &   6.150     \\
           7$D_{5/2}$ &   26068.77     &   2.890      \\
           8$D_{5/2}$ &   27822.88     &   1.810      \\
           9$D_{5/2}$ &   28835.79     &   1.169      \\
          10$D_{5/2}$ &   29472.94     &  0.909      \\
          11$D_{5/2}$ &   29899.55     &  0.735      \\
          12$D_{5/2}$ &   30199.10     &  0.630      \\
          13$D_{5/2}$ &   30417.46     &  0.642      \\
          14$D_{5/2}$ &   30581.53     &  0.699      \\
          15$D_{5/2}$ &   30707.91     &  0.741      \\
          16$D_{5/2}$ &   30807.33     &  0.766      \\
          17$D_{5/2}$ &   30886.94     &  0.798      \\
          18$D_{5/2}$ &   30951.68     &  0.745      \\
          19$D_{5/2}$ &   31005.03     &  0.903      \\
          20$D_{5/2}$ &   31049.52     &  0.840      \\
          21$D_{5/2}$ &   31087.00     &   1.438      \\
          22$D_{5/2}$ &   31118.87     &  0.130      \\
          23$D_{5/2}$ &   31146.20     &   1.456      \\
          24$D_{5/2}$ &   31169.81     &   1.288      \\
          25$D_{5/2}$ &   31190.35     &  0.149      \\
          26$D_{5/2}$ &   54508.05     &  0.998      \\
          27$D_{5/2}$ &   69450.32     &  0.713      \\
          28$D_{5/2}$ &   92156.01     &  0.462      \\
          29$D_{5/2}$ &   126462.2     &  0.280      \\
          30$D_{5/2}$ &   178123.1     &  0.158      \\
          31$D_{5/2}$ &   255714.2     &  0.081 \\
          32$D_{5/2}$ &   372021.2     &  0.035 \\
          33$D_{5/2}$ &   546142.6     &  0.008 \\
          34$D_{5/2}$ &   806627.1     &  0.005 \\
          35$D_{5/2}$ &   1196092      &  0.010 \\
          36$D_{5/2}$ &   1778024      &  0.011 \\
          37$D_{5/2}$ &   2646960      &  0.010 \\
          38$D_{5/2}$ &   3944018      &  0.008 \\
          39$D_{5/2}$ &   5880432      &  0.006 \\
          40$D_{5/2}$ &   8771401      &  0.005 \\
          41$D_{5/2}$ &  13082459      &  0.003 \\
          42$D_{5/2}$ &  19494500      &  0.002 \\

\noalign{\smallskip}\hline
\end{tabular}
\end{table}

\end{document}